\documentclass[fleqn,usenatbib]{mnras}

\usepackage{newtxtext,newtxmath}
\usepackage{natbib}
\usepackage{multicol}
\usepackage{xcolor}

\usepackage[T1]{fontenc}
\usepackage{ae,aecompl}

\usepackage{graphicx}	
\usepackage{amsmath}	
\usepackage{amssymb}	
\usepackage{bm}

\usepackage{comment}

\newcommand{\MHI}{M_\mathrm{HI}}
\newcommand{\rstarhalf}{R_{*,1/2}}
\newcommand{\dlim}{d_\mathrm{host}\leq300\mathrm{\,kpc}}

\newcommand{\mstarlowlim}{M_*\geq10^5\mathrm{\,\Msun}}

\newcommand{\mystar}{*}
\newcommand{\Msun}{\mathrm{M}_{\odot}}
\newcommand{\Mstar}{M_{\mystar}}
\newcommand{\Mgas}{M_\mathrm{gas}}

\newcommand{\Mtwohm}{M_\mathrm{200m}}
\newcommand{\Mpeak}{M_\mathrm{peak}}
\newcommand{\lcdm}{\Lambda\mathrm{CDM}}

\newcommand{\rtwoh}{R_{\rm 200}}
\newcommand{\rtwohm}{R_{\rm 200m}}

\newcommand{\vmax}{V_\mathrm{max}}

\newcommand{\mpc}{\mathrm{Mpc}}

\newcommand{\kpc}{\mathrm{kpc}}

\newcommand{\gyr}{\mathrm{Gyr}}
\newcommand{\myr}{\mathrm{Myr}}

\newcommand{\dhost}{d_{\rm host}}

\title[Extinguishing the FIRE]{Extinguishing the FIRE: environmental quenching of satellite galaxies around Milky Way-mass hosts in simulations}

\author[J. Samuel et al.]
{Jenna Samuel$^{1,2}$\thanks{E-mail: jenna.samuel@austin.utexas.edu}\thanks{NSF Astronomy and Astrophysics Postdoctoral Fellow},
Andrew Wetzel$^{2}$,
Isaiah Santistevan$^{2}$,
Erik Tollerud$^{3}$,
\newauthor
Jorge Moreno$^{4,5}$,
Michael Boylan-Kolchin$^{1}$,
Jeremy Bailin$^{6}$,
Bhavya Pardasani$^{7}$
\\
$^{1}$Department of Astronomy, The University of Texas at Austin, 2515 Speedway, Stop C1400, Austin, TX 78712, USA\\
$^{2}$Department of Physics and Astronomy, University of California, Davis, CA 95616, USA\\
$^{3}$Space Telescope Science Institute, 3700 San Martin Dr, Baltimore, MD 21218, USA\\
$^{4}$Department of Physics and Astronomy, Pomona College, Claremont, CA 91711, USA\\
$^{5}$Downing College, University of Cambridge, Cambridge CB3 OHA, UK\\
$^{6}$Department of Physics and Astronomy, University of Alabama, Box 870324, Tuscaloosa, AL 35487-0324, USA\\
$^{7}$Department of Astronomy, University of Illinois at Urbana-Champaign, 1002 W Green St, Urbana, IL 61801, USA
}

\pubyear{2022}

\begin{document}
\label{firstpage}
\pagerange{\pageref{firstpage}--\pageref{lastpage}}
\maketitle

\begin{abstract}
The star formation and gas content of satellite galaxies around the Milky Way (MW) and Andromeda (M31) are depleted relative to more isolated galaxies in the Local Group (LG) at fixed stellar mass.
We explore the environmental regulation of gas content and quenching of star formation in $z=0$ galaxies at $M_{*}=10^{5-10}\,\rm{M}_{\odot}$ around 14 MW-mass hosts from the FIRE-2 simulations.
Lower-mass satellites ($M_{*}\lesssim10^7\,\rm{M}_{\odot}$) are mostly quiescent and higher-mass satellites ($M_{*}\gtrsim10^8\,\rm{M}_{\odot}$) are mostly star-forming, with intermediate-mass satellites ($M_{*}\approx10^{7-8}\,\rm{M}_{\odot}$) split roughly equally between quiescent and star-forming.
Hosts with more gas in their circumgalactic medium have a higher quiescent fraction of massive satellites ($M_{*}=10^{8-9}\,\rm{M}_{\odot}$).
We find no significant dependence on isolated versus paired (LG-like) host environments, and the quiescent fractions of satellites around MW-mass and LMC-mass hosts from the FIRE-2 simulations are remarkably similar.
Environmental effects that lead to quenching can also occur as preprocessing in low-mass groups prior to MW infall.
Lower-mass satellites typically quenched before MW infall as central galaxies or rapidly during infall into a low-mass group or a MW-mass galaxy.
Most intermediate- to high-mass quiescent satellites have experienced $\geq1-2$ pericentre passages ($\approx2.5-5$ Gyr) within a MW-mass halo.
Most galaxies with $M_{*}\gtrsim10^{6.5}\,\rm{M}_{\odot}$ did not quench before falling into a host, indicating a possible upper mass limit for isolated quenching.
The simulations reproduce the average trend in the LG quiescent fraction across the full range of satellite stellar masses. 
Though the simulations are consistent with the SAGA survey’s quiescent fraction at $M_{*}\gtrsim10^8\,\rm{M}_{\odot}$, they do not generally reproduce SAGA’s turnover at lower masses.
\end{abstract}



\section{Introduction}

Low-mass galaxies ($\Mstar\lesssim10^9\,\Msun$) are especially susceptible to the disruptive effects of stellar feedback and the environment because of their shallow gravitational potentials.
At very low mass ($\Mstar\lesssim10^5\,\Msun$), ultra-faint galaxies are likely quenched by reionization because they are unable to shield their gas from the dissociative effects of ultraviolet (UV) photons in the early Universe \citep{Bullock2000,Benson2002,Weisz2014b,Fitts2017,RodriguezWimberly2019}.
In simulations, the cold gas that fuels star formation can be repeatedly heated and ejected from low-mass galaxies by their bursty episodes of star formation and the ensuing feedback from supernovae \citep[e.g.,][]{ElBadry2018}.
The ejected gas generally cools and re-accretes into isolated low-mass galaxies for further star formation, explaining why observed isolated galaxies at $\Mstar=10^{7-9}\,\Msun$ remain star-forming \citep{Geha2012}.

Satellite galaxies are subjected to additional disruptive processes within the host environment that isolated galaxies are not exposed to.
For example, the relative motion of satellites through the circumgalactic medium (CGM) of the host causes satellites to experience ram pressure \citep[e.g.,][]{Gunn1972,McCarthy2008,Grcevich2009,Simons2020}, which may act together with stellar feedback to more efficiently remove gas from satellites and quench their star formation.
Thus, satellite galaxies are more likely to be quenched than their isolated counterparts at fixed stellar mass.

Satellite galaxies in the Local Group (LG) are almost all gas-poor and quiescent, except for some of the most massive satellites like the Magellanic Clouds, LGS3, and IC10 \citep[e.g.,][]{McConnachie2012,Spekkens2014,Wetzel2015b}. 
In fact, \citet{Tollerud2018} showed that the lack of faint LG dwarf galaxies containing HI is so severe that it requires reionization in addition to environmental effects to be explainable in a $\Lambda$CDM context.
\citet{Putman2021} presented an updated compilation of neutral hydrogen (HI) measurements in LG galaxies, revealing that they are increasingly HI-poor as both distance from the closest massive galaxy (either the MW or M31) and distance from the surface of the LG decreases.
This could mean that the paired nature of the MW and M31 has an additional disruptive effect on satellite galaxies compared to an isolated MW-like environment, possibly from a more massive or dense CGM encompassing both hosts.

In contrast to the LG, results from the Satellites Around Galactic Analogs (SAGA) survey indicate that most satellite galaxies (M$_r < -12.3$; $\Mstar\gtrsim5\times10^6\,\Msun$) of nearby isolated MW analogues are still star-forming at $z\approx0$ \citep{Geha2017,Mao2021}.
This difference in the quiescent fraction of satellites presents a unique tension, given that the host galaxies in SAGA are selected specifically to match the MW's stellar mass, and the number and radial distribution of satellites around SAGA hosts broadly agrees with LG satellites and zoom-in simulations of MW-mass galaxies \citep[e.g.,][]{Samuel2020,Font2021}.
Furthermore, \citet{Font2022} used the colors of satellites of MW-mass galaxies in the local Universe, which have been surveyed extensively \citep[e.g.,][]{Karachentsev2014,Muller2015,Danieli2017,Smercina2018,Bennet2019,Crnojevic2019,Muller2019,Carlsten2021,Carlsten2022}, to estimate their star-forming or quenched status and found broad agreement between the LG and the local Universe, extending the observational tension between the SAGA survey and and other MW-mass galaxies.

This tension could originate from key physical differences in host environments in the LG compared to SAGA hosts; alternatively, there could be distinct observational selection functions in each data set that lead to different measured quiescent fractions.
\citet{Karunakaran2021} repeated the SAGA quiescent fraction analysis with archival GALEX UV imaging and compared this to the APOSTLE and Auriga simulations \citep{Fattahi2016,Sawala2016,Grand2017}.
The results of their observational analysis support the original SAGA quiescent fraction, but their simulation results are consistent with the LG, leading them to claim that there may be a physical difference in satellite evolution around SAGA hosts.
Using the ARTEMIS simulations, \citet{Font2022} showed that the inclusion of an observational selection function, or correction for the detectability of galaxies based on their surface brightness, has a significant effect on the measured quiescent fraction, which can account for differences between the LG and the SAGA survey.
Moreover, the SAGA survey could also be including interlopers or non-satellites because of projection effects along the line of sight.
However, it is unclear if there is a single cause of the tension between LG and SAGA satellites.

The physical processes involved in quenching low-mass satellite galaxies and the timescales on which they act have been widely studied in simulations.
\citet{Simpson2018} found significant environmental quenching from ram pressure for satellite galaxies ($\Mstar<10^7\,\Msun$) using the Auriga simulations, which is broadly consistent with the LG.
Similarly, \citet{Buck2019} found that satellites in the NIHAO simulations have low gas fractions compared to isolated galaxies because of enhanced ram pressure during pericentre passages in the host halo.
Using a large sample of satellite galaxies from the TNG50 simulations, \citet{Joshi2021} demonstrated that satellites form the bulk of their stellar mass prior to infall into a massive host or cluster environment.
In addition, \citet{Fillingham2015} and \citet{Wetzel2015b} found that environmental quenching proceeds on short timescales of about 2 Gyr for low- to intermediate-mass satellites, and that complete gas consumption on longer timescales of about 8 Gyr is the main mode of quenching for more massive satellites.
\citet{Akins2021} also showed a similar positive trend in quenching timescales versus stellar mass using the ChaNGa DC Justice League suite of simulations.

However, satellites may quench even before entering a MW-mass halo through processes such as reionization, internal stellar feedback, and environmental quenching in lower-mass groups (group preprocessing), in particular.
\citet{Wetzel2015a} argued that a significant fraction of satellite galaxies in MW-mass haloes could be preprocessed in lower-mass groups, and the same environmental effects that act to quench satellites in a MW-like environment also can manifest in lower-mass groups.
\citet{Jahn2021} demonstrated that effective environmental quenching of satellite galaxies occurs around isolated LMC-mass hosts from the FIRE-2 simulations, and there is significant evidence that the LMC has brought its own satellite population into the MW's halo \citep{Li2008,DOnghia2008,Sales2017,Kallivayalil2018,Patel2020,SantosSantos2021}.
Such preprocessing effects might augment environmental quenching in MW-mass haloes for satellites that were previously in lower-mass groups, so it is important to disentangle these distinct quenching phases.

We present an overview of the quiescent fraction of $z=0$ satellite galaxies around MW-mass hosts in the FIRE-2 simulations.
We further investigate when and where they quenched.
In Section~\ref{sec:simulations} we discuss the details of our simulations, how we assigned gas cells to galaxies, and how we define quiescence.
We present our main results in Section~\ref{sec:results}, including comparisons of the quiescent fraction of simulated satellites to observations, trends in quiescent fraction with respect to host environment and group preprocessing, quenching timescales relative to infall and pericentre passages, and three case studies of satellite quenching.
We summarize our work and discuss its implications in Section~\ref{sec:discussion}.


\section{Simulations}\label{sec:simulations}

We use the Latte \citep{Wetzel2016} and ELVIS on FIRE \citep{GK2019a,GK2019b} suites from the FIRE-2 cosmological baryonic simulations \citep{Hopkins2018} of the Feedback In Realistic Environments (FIRE) project\footnote{\url{https://fire.northwestern.edu/}}.
These simulation suites model isolated and LG-like pairs of 14 MW/M31-mass host galaxies.
The six paired hosts (ELVIS) and one isolated host (m12z) were simulated with m$_{\rm baryon,ini}=3500-4200\,\Msun$ (m$_{\rm dm}\approx2\times10^4\,\Msun$), and the other seven isolated hosts (Latte) have m$_{\rm baryon,ini}=7100\,\Msun$ (m$_{\rm dm}=3.5\times10^4\,\Msun$). 
The hosts have total masses $M_{\rm 200m}=1-2\times10^{12}\,\Msun$ and disc stellar masses $\Mstar\approx10^{10-11}\,\Msun$, which are within observational uncertainties of the MW's properties\footnote{`200m' indicates a measurement relative to 200 times the mean matter density of the Universe.}.
These simulations reproduce the stellar mass functions, radial distributions, star-formation histories, and some aspects of satellite planes in the LG \citep{Wetzel2016,GK2019a,GK2019b,Samuel2020,Samuel2021}.

We ran the simulations with the FIRE-2 implementations of fluid dynamics, star formation, and stellar feedback \citep{Hopkins2018}.
FIRE uses the \textsc{GIZMO} Lagrangian meshless finite-mass (MFM) hydrodynamics code \citep{Hopkins2015}.
\textsc{GIZMO} enables adaptive hydrodynamic gas smoothing based on the local density of gas cells while still conserving mass, energy, and momentum to machine accuracy. 
The gas gravitational softening is fully adaptive, matched to the hydrodynamic resolution (inter-cell spacing), and the minimum softening length reached is $\approx 1$ pc.
Gravitational forces are solved using an upgraded version of the $N$-body \textsc{GADGET-3} Tree-PM solver \citep{Springel2005}.

The FIRE-2 methodology includes detailed subgrid models for gas physics, star formation, and stellar feedback.
Gas models used include: a metallicity-dependent treatment of radiative heating and cooling over $10-10^{10}$ K \citep{Hopkins2018}, a cosmic ultraviolet background with early HI reionization ($z_{\rm reion}\sim10$) \citep{FaucherGiguere2009}, and turbulent diffusion of metals in gas \citep{Hopkins2016,Su2017,Escala2018}.
Star formation occurs in gas that is self-gravitating, Jeans-unstable, cold (T $<10^4$ K), dense ($n>1000$ cm$^{-3}$), and molecular (following \citealt{Krumholz2011}).
We model several stellar-feedback processes, including core-collapse and Ia supernovae, continuous stellar mass loss, photoionization, photoelectric heating, and radiation pressure.

We generated cosmological zoom-in initial conditions at $z \approx 99$ using \textsc{MUSIC} \citep{Hahn2011}.
All simulations assume flat $\lcdm$ cosmologies, with slightly different parameters across the full suite: $h = 0.68 - 0.71$, $\Omega_\Lambda = 0.69 - 0.734$, $\Omega_m = 0.266 - 0.31$, $\Omega_b = 0.0455 - 0.048$, $\sigma_8 = 0.801 - 0.82$, and $n_{\rm s} = 0.961 - 0.97$, broadly consistent with \citet{PlanckCollaboration2018}.

\begin{figure}
	\includegraphics[width=\columnwidth]{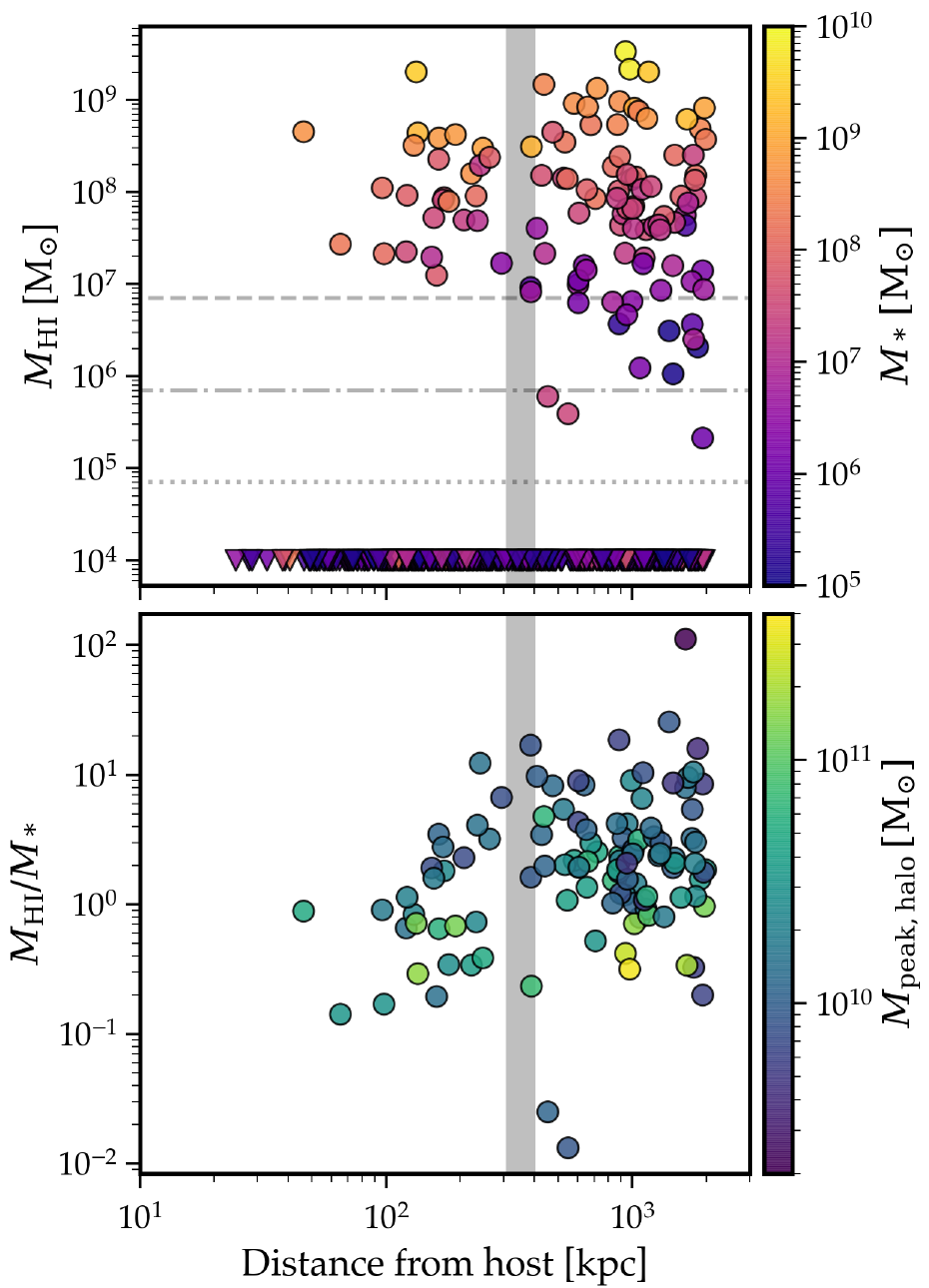}\par
	\vspace{-2 mm}
    \caption{Gas content of the (529) galaxies ($\Mstar=10^{5-10}\,\Msun$) out to 2 Mpc from the MW-mass hosts at $z=0$.
    \textit{Top:} HI gas mass versus distance from the nearest MW-mass host. 
    We show galaxies with $\MHI<10^4\, \Msun$ (our resolution limit) as triangles at the bottom.
    The horizontal lines show the mass of 10, 100, and 1000 gas cells, and the vertical shaded region shows the range of host radii ($R_{\rm 200m}$).
    Most galaxies have little to no gas, and there is a noticeable lack of galaxies with $\MHI\lesssim10^6\,\Msun$ ($\approx140$ gas cells) at all distances, revealing a bimodal nature of gas content similar to LG galaxies.
    \textit{Bottom:} The ratio of HI gas mass to stellar mass versus distance from the nearest MW-mass host for the 112 galaxies with $\MHI$ above our resolution limit.
    Within the host radii, this ratio decreases as distance from the host decreases.
    These are likely satellites that are undergoing environmental quenching processes, but have not yet quenched.
    }
    \label{fig:HI_content}
\end{figure}

\subsection{Gas assignment}\label{gas_assign}

Using the \textsc{ROCKSTAR} 6D halo finder \citep{Behroozi2013a}, we identify dark matter (sub)haloes in our simulations at each of the 600 snapshots per simulation and construct merger trees using \textsc{CONSISTENT-TREES} \citep{Behroozi2013b}. 
We assign star particles to dark matter haloes in a post-processing step described further in \citet{Necib2019} and \citet{Samuel2020}.

We assign gas cells to subhaloes at $z=0$ in a manner similar to star particle assignment.
We use a subhalo catalog to identify the center position, bulk velocity, and other properties of each subhalo.
We require the gas cells to be (1) within twice the stellar half mass radius ($\rstarhalf$) of a subhalo and (2) within twice the greater of either the subhalo maximum circular velocity ($\vmax$) or the velocity dispersion ($\sigma$) of DM particles belonging to the subhalo.
To determine a satellite's $\MHI$, we use the stored HI fraction for each of its assigned gas cells, which is calculated according to the collisional ionization of gas (from its density, temperature, pressure, and assuming local thermodynamic/ionization equilibrium) and from the incident ionizing radiation \citep{Hopkins2018}.
We tested different distance criteria and found that for satellites containing HI inside of $\rstarhalf$ the amount of HI enclosed rose steadily from 1-2 $\rstarhalf$ and plateaued from 2-4 $\rstarhalf$, whereas satellites with no HI inside of $\rstarhalf$ began to pick up spurious HI between 2-4 $\rstarhalf$.
Using up to a factor of 10 times $\vmax$ or $\sigma$ at fixed radius resulted in no significant changes to the HI mass assigned to subhaloes compared to our nominal factor of 2.

In Figure~\ref{fig:HI_content} (top) we show the HI gas mass of the 529 galaxies within the zoom-in region of our simulations, a $\approx2\,\mpc$ radius around each MW host.
We select galaxies that have stellar masses $\Mstar=10^{5-10}\,\Msun$ and reside in well-resolved dark matter subhaloes \citep{Samuel2020}.
Nearly all of the gas-rich lower-mass ($\Mstar\lesssim10^7\,\Msun$) galaxies exist outside of the host virial radius ($\approx300-400$ kpc), indicating that strong environmental effects likely remove gas from lower-mass satellites ($\dlim$).
Only 26 satellites contain HI gas, and all of those have $\Mstar\gtrsim10^7\,\Msun$, consistent with observations of HI in LG galaxies where only the most massive satellites have detectable HI \citep{Spekkens2014,Putman2021}.
In Figure~\ref{fig:HI_content} (bottom) we further demonstrate that the (112) galaxies with HI gas display a decrease in the ratio of $\MHI/\Mstar$ as distance from the MW-mass host decreases, relative to galaxies outside the MW-mass halos.
This most likely reflects that these satellites are in the process of quenching within the MW-mass host environment, and that satellites at smaller distances are experiencing stronger environmental quenching like enhanced ram pressure at smaller distances where the host's CGM density is higher.

The galaxies that contain HI gas above our resolution limit ($\approx10^4\,\Msun$) typically have significant amounts of gas that are well-resolved with over 100 gas cells ($\gtrsim10^6\,\Msun$).
\citet{Rey2022} found a similar bimodality in the gas content of isolated galaxies ($\Mstar\approx10^{4-6}\,\Msun$) in simulations because of a variety of formation histories. 
However, in the following sections we examine both environmental and numerical effects, which are the likely drivers of the separation of our galaxies into HI-rich and HI-poor.

\subsection{Definition of quiescence}\label{sec:q_def}

Before examining environmental quenching, we must define what it means for a galaxy to be quiescent in our simulations.
Common observational definitions of quiescence use star formation rate (SFR) indicators like the presence of detectable HI or H$\alpha$ \citep[e.g.,][]{Grcevich2009,Spekkens2014,Putman2021}.
The full time-resolved star formation history (SFH) and $\MHI$ of each galaxy are readily computed from the simulation output, so we use these quantities to build a definition of quiescence.

Galaxies often remain star-forming in isolated environments, so we first determine how long on average it would take a lower-mass ($\Mstar=10^{5-6}\,\Msun$), isolated, star-forming galaxy to form one star particle worth of stellar mass ($m_{\rm{baryon}}=3500-7070\,\Msun$).
If a galaxy has not had any star formation in this length of time, we might reasonably label it as quiescent because higher-mass galaxies should form stars quicker. 
A galaxy that has not formed stars in this length of time could potentially form stars at some later time if it still contains gas, but it is unlikely to have any observable traces of recent star formation like H$\alpha$, which probes timescales as short as $5\,\myr$ in the FIRE simulations \citep{Flores2021}.
\citet{Fitts2017} studied the SFRs of isolated galaxies ($\Mstar\approx10^{5-7}\,\Msun$) in simulations using the same FIRE-2 physics as the simulations used here, but at 14 times better resolution ($m_{\rm{baryon}}=500\,\Msun$).
They found that these galaxies experience bursts of star formation on $50\,\myr$ timescales and generally continue to be star-forming to $z=0$.
They also found that average SFR correlates with galaxy stellar mass; a galaxy with $\Mstar=4.7\times10^5\,\Msun$ had an average SFR of $3.5\times10^{-5}\,\Msun\rm{yr}^{-1}$ over all of cosmic time, and a galaxy with $\Mstar=4.1\times10^6\,\Msun$ had a SFR of $3\times10^{-4}\,\Msun\rm{yr}^{-1}$.
Extrapolating these SFRs to the simulations we use in this work, it should take similarly massive galaxies $\approx200\,\myr$ and $\approx25\,\myr$ to form one star particle ($7070\,\Msun$), with the more massive galaxy being able to form a star particle in less time.

In Appendix~\ref{sec:appendix_q_def} we show the effects of varying the lookback time to last star formation, demonstrating that using $100-600\,\myr$ yields relatively similar quiescent fractions for satellite galaxies ($\dlim$).
Unfortunately, using only lookback time to last star formation as a quiescence definition yields some other undesirable effects: our results for subsamples of the simulations (separating satellites by host or prior low-mass group association) can be sensitive to the exact time chosen, and when we use a lookback time of $200\,\myr$ we label six satellites that contain significant HI reservoirs ($\MHI\approx10^{7-8}\,\Msun$) as quiescent.

In Section~\ref{sec:case_studies}, we observe in a few examples that star formation usually occurs when a galaxy has a well-resolved amount of HI ($\MHI\gtrsim10^6\,\Msun$, or $\gtrsim140$ gas cells).
In Appendix~\ref{sec:appendix_q_def}, we also show the quiescent fraction of satellites using alternate definitions of quiescence based solely on the absence of HI gas ($\MHI<10^{4-7}\,\Msun$).
While the quiescent fraction that we measure is qualitatively similar to the lookback time criterion, we unsuitably label satellites that have recently exhausted their gas supply, but formed stars within the last few snapshots of the simulation, as quiescent.
Moreover, \citet{Rey2022} found that the HI mass and its spatial extent in simulations of isolated low-mass galaxies ($\Mstar\approx10^{4-6}\,\Msun$) vary significantly over the last few Gyr because of stellar feedback, so the HI assigned to our galaxies at $z=0$ alone is unlikely to give a definitive answer on quiescence.

However, when we combine criteria for lookback time to last star formation and $\MHI$ at $z=0$, we find that it yields a conservative and robust theoretical definition of quiescence. 
A quiescent galaxy under a combined definition has had no recent star formation and likely does not contain sufficient fuel for future star formation. 
In addition, our subsampled results become insensitive to the specific lookback time used in the definition of quiescence when we combine it with an $\MHI$ criterion.
We thus define a galaxy as quiescent if no stars have formed within it for the last $200\,\myr$ \textit{and} it has $\MHI<10^6\,\Msun$ ($\lesssim140$ gas cells) at $z=0$.

Notably, using this definition we would classify some observed transitional LG galaxies such as Leo T, Phoenix, and LGS 3 as quiescent, which are typically considered star-forming on a low level, because they have $\MHI<10^6\,\Msun$ and formed their last stars $\approx500-1000\,\myr$ ago \citep{Weisz2014a}.
However, we would classify other LG galaxies that are also considered star-forming on a low level (such as Pegasus and DDO 210) as star-forming because they have both formed stars within the last $200\,\myr$ and contain $\MHI\gtrsim10^6\,\Msun$.
We have verified that adjusting our definition of quiescence to classify Leo T, Phoenix, and LGS 3 as star-forming by changing the threshold for HI gas to $\MHI<10^5\,\Msun$ and/or changing the lookback time to last star formation to $500\,\myr$ does not qualitatively change our conclusions. 
From the previous section, this is because our simulated galaxies typically have either no HI or $\MHI\gtrsim10^6\,\Msun$, likely because the simulations may numerically over-quench satellites at low masses ($\Mstar=10^{5-6}\,\Msun$), as we discuss in Section~\ref{sec:QF}.

To characterize statistical uncertainties on measured quiescent fractions for different galaxy samples from the simulations, we estimate 1-sigma uncertainties using a Bayesian method that adopts a uniform prior for the true quiescent fraction.
We assume that the data in each stellar mass bin are drawn from a beta continuous random variable distribution.
We measure the lower and upper 1-sigma uncertainties using SciPy \citep{scipy} by calculating the 16th and 84th quantiles of a beta distribution with shape parameters $a=N_{\rm quench}+1$ and $b=N-N_{\rm quench}+1$, where $N$ is the total number of galaxies in a bin.
\citet{Cameron2011} showed that this method returns 1-sigma confidence intervals (CIs) that contain the true population proportion (e.g., quiescent fraction) $\approx70$ per cent of the time for sample sizes $N=1-100$, and the method avoids the under-estimation and over-estimation of CIs of other common estimators at low $N$.


\section{Results}\label{sec:results}

Throughout this work, unless explicitly stated otherwise, we define quiescence as the combination of no star formation within a galaxy in the last 200 Myr and $\MHI<10^6\,\Msun$ at $z=0$ and we show statistical estimates of 1-sigma uncertainties on quiescent fractions that combine galaxies across simulations (see Section~\ref{sec:q_def} for details).
Sections~\ref{sec:qf_obs} and~\ref{sec:qf_host_env} are the only exceptions where we show host-to-host scatter instead of statistical uncertainty, and we implement different definitions of quiescence for the mock surveys in Section~\ref{sec:qf_obs}.

\begin{figure}
	\includegraphics[width=0.47\textwidth]{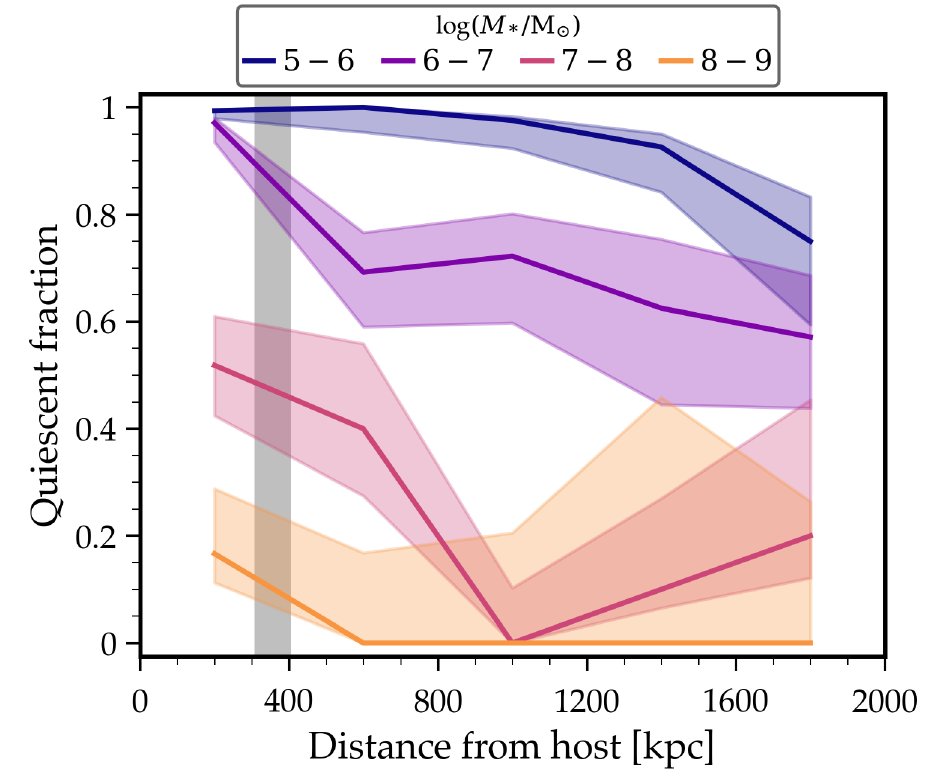}
    \vspace{-3 mm}
    \caption{The quiescent fraction of galaxies versus distance from the nearest MW-mass host out to 2 Mpc.
    The uncertainties on the data are statistical estimates described at the top of Section~\ref{sec:results}.
    The vertical grey shaded region shows the range of MW-mass host radii ($R_{\rm 200m}$).
    Most galaxies with $\Mstar=10^{5-6}\,\Msun$ are quiescent throughout the zoom-in regions, and we do not show the (8) galaxies with $\Mstar=10^{9-10}\,\Msun$ because they are all star-forming.
    The galaxies in between these two extremes exhibit a noticeable increase in quiescent fraction closer to the MW-mass host, demonstrating the importance of the host environment in their quenching.
    }
    \label{fig:qf_vs_dist}
\end{figure}

In Figure~\ref{fig:qf_vs_dist}, we show the quiescent fraction of all galaxies versus distance from the nearest MW-mass host out to 2 Mpc.
The quiescent fraction for galaxies with $\Mstar=10^{5-9}\,\Msun$ generally increases as distance from the host decreases.
The noticeable break in this trend for the star-forming satellites with $\Mstar=10^{7-8}\,\Msun$ at distances of $800-1200\,\mpc$ is largely stochastic.
There is only one quiescent galaxy per distance bin in this mass range beyond 1200 kpc, but the number of galaxies in each bin decreases with distance, so the quiescent fraction increases.
The lowest-mass galaxies ($\Mstar=10^{5-6}\,\Msun$) are almost all quiescent except at large distances ($1-2\,\mpc$) where the quiescent fraction drops to a minimum of $\approx75$ per cent.
The higher-mass galaxies ($\Mstar=10^{8-9}\,\Msun$) are all star-forming except for about 20 per cent of them within $400\,\kpc$.
The eight galaxies with $\Mstar>10^{9}\,\Msun$ are all star-forming, so we omit them from this figure.
The fact that the population of galaxies at $\Mstar<10^7\,\Msun$ is over 50 per cent quiescent at all distances could indicate far-reaching effects of the host environment or over-quenching from finite numerical resolution \citep{Hopkins2018}, which we discuss in the following section.

\begin{figure}
	\includegraphics[width=0.47\textwidth]{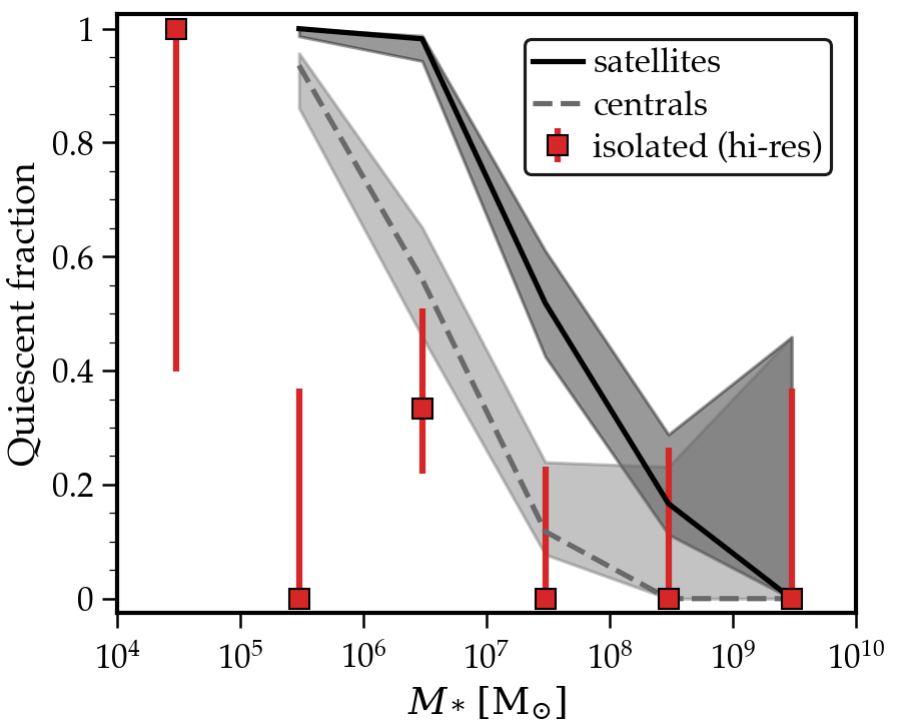}
    \vspace{-2 mm}
    \caption{A convergence test of the numerical resolution in the MW simulations.
    We show the quiescent fraction versus stellar mass of galaxies in three samples: satellites and centrals from the MW simulations, and isolated galaxies from higher-resolution simulations. 
    Satellites are within 300 kpc of the nearest MW-mass host, and centrals are $1-2$ Mpc from the MW-mass host and have never been within $\rtwohm$ of a more massive halo.
    The higher-resolution isolated galaxies are the primary haloes in those simulations and there are no MW-mass haloes within their zoom-in regions.
    Satellites are more likely to be quiescent than centrals at fixed resolution, as expected from environmental quenching within a MW-mass halo.
    The reasonable agreement between centrals and higher-resolution isolated galaxies at $\Mstar\gtrsim10^6\,\Msun$ implies that our results are numerically well-converged there, but the difference between them at $\Mstar=10^{5-6}\,\Msun$ could indicate numerical over-quenching at such low masses in the (lower-resolution) MW simulations.
    }
    \label{fig:qf_isolated}
\end{figure}

\begin{figure*}
    \begin{multicols}{2}
	\includegraphics[width=0.47\textwidth]{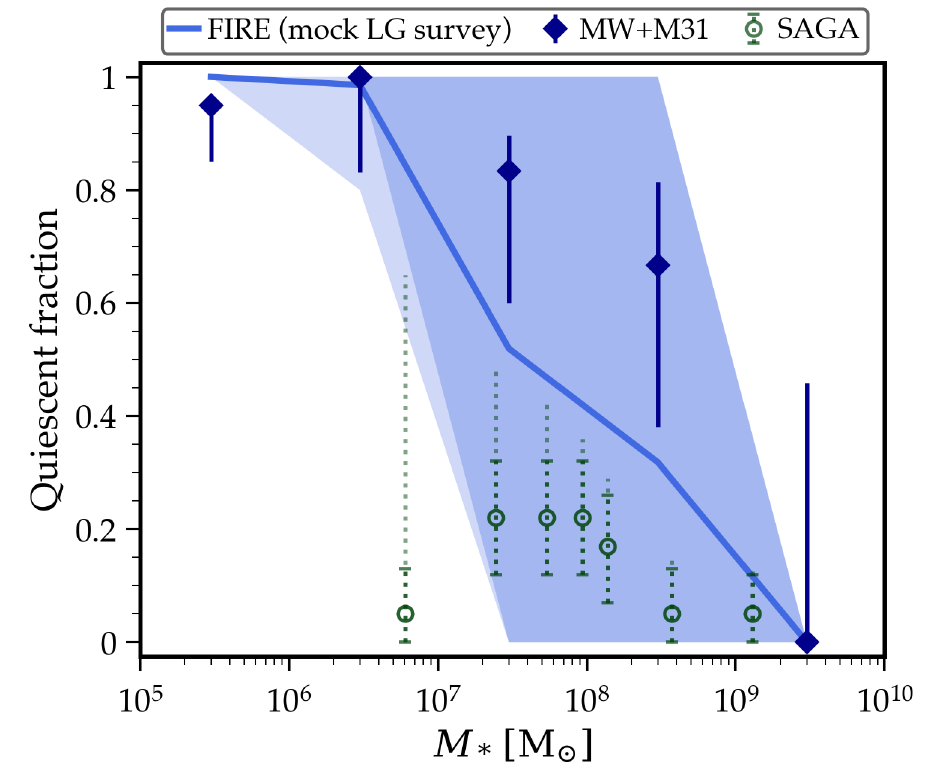}\par
	\includegraphics[width=0.47\textwidth]{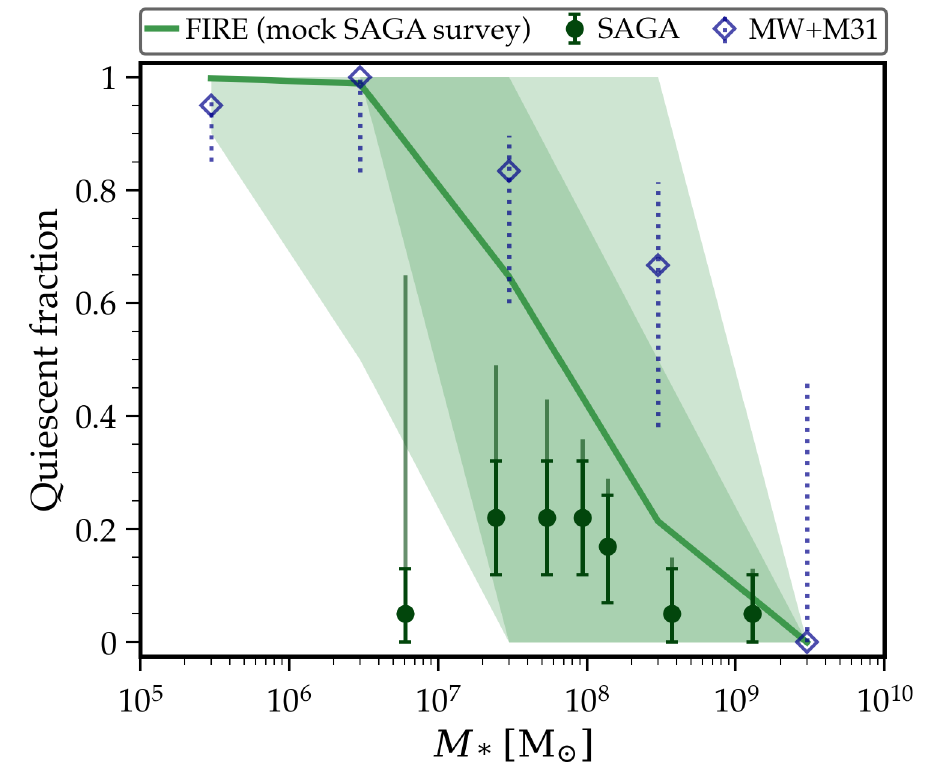}\par
	\end{multicols}
    \vspace{-8 mm}
    \caption{Comparisons of the quiescent fraction of satellites in the simulations and observations of the LG (left) and the SAGA survey (right).
    The solid lines and shaded regions are the mean and 68 (100) per cent host-to-host scatter across the 14 MW-mass hosts from the FIRE-2 simulations.
    Unfilled symbols and dotted lines in each panel are not a direct comparison to the simulations, but we show them as a reference.
    \textit{Left:} 
    We use $\MHI<10^6\,\Msun$ to define quiescence here, similar to the HI detection criteria for LG observations.
    We have updated the LG results from \citet{Wetzel2015b} using data from \citet{Putman2021}.
    Simulated satellites with $\Mstar\lesssim10^7\,\Msun$ are almost all quiescent, similar to the LG.
    At $\Mstar>10^{7}\,\Msun$, the simulations are broadly consistent with the LG at the 1-sigma level, because of the large host-to-host scatter.
    \textit{Right:} We select simulated satellites in 2D projection to mimic the selection function in SAGA, which includes potential contamination from line-of-sight non-satellites (see text for details).
    We define quiescence here via the lack of star formation within the last 200 Myr, motivated by SAGA's H$\alpha$-based definition that probes short timescales.
    We take SAGA results from \citet{Mao2021}.
    Compared to the left panel, projection effects shift the lower bound on the simulation distribution down, towards the lowest-mass and highly star-forming bin in SAGA.
    The simulation mean also shifts down for $\Mstar>10^{8}\,\Msun$, closer to the SAGA results.
    In each panel, the large host-to-host scatter leads to broad consistency with both the LG and SAGA.
    }
    \label{fig:qf_obs}
\end{figure*}

\subsection{Satellite quiescent fraction}\label{sec:QF}

We primarily focus our analysis on the satellite galaxies ($\dlim$, comparable to the smallest $\rtwohm$ of the hosts) at $z=0$ to examine the effects of the MW-mass host environment on satellite quenching.
We show the stellar mass distribution of our satellites in Appendix~\ref{sec:appendix_QDT}.
By our fiducial definition of quiescence, 209 of the 240 total satellite galaxies are quiescent.
Several higher-mass satellites are gas-rich and star-forming, but satellites with $\Mstar\lesssim10^7\,\Msun$ are almost all quiescent.
The lack of low-to-intermediate mass satellite galaxies containing any gas, and the presence of star-forming and gas-rich galaxies outside the host halo radius suggests effective environmental gas removal and quenching within the host haloes.
This is consistent with \citet{GK2019b}, who found that the galaxies in the outskirts of these simulations (non-satellites) have more extended star formation histories than satellites, and environmental effects seem to extend farther from the paired LG-like hosts than the isolated MW-mass hosts.
See \citet{GK2019b} for more details on the star formation histories of the galaxies in these simulations.

\subsubsection{Numerical resolution}

A large total quiescent fraction of satellites could potentially be caused by numerical rather than physical effects.
If the resolution of a simulation is too low, gas could be unphysically removed from (especially lower-mass) galaxies causing numerical over-quenching.
To assess possible resolution effects in the MW simulations, we compare the quiescent fraction of simulated galaxies in and around the MW-mass haloes with 27 higher-resolution isolated galaxies from the FIRE-2 simulations \citep{Fitts2017,ElBadry2018,Hopkins2018,Chan2018a,Wheeler2019}.
The high-resolution isolated sample is comprised of zoom-in simulations of galaxies ($\Mstar\approx10^{4-9}\,\Msun$ and $\Mtwohm\approx10^{9-11}\,\Msun$ at $z=0$) with no more-massive galaxy in the zoom-in region, so they are not environmentally quenched by a more massive host or nearby galaxy.
These isolated simulations exhibit a variety of star formation histories and gas morphologies consistent with observations of isolated galaxies of similar mass.
All isolated simulations were run with the same reionization model as the MW-mass simulations used in this work.
Most of the isolated simulations are higher resolution than the MW simulations, with baryonic mass resolutions of typically $m_{\rm baryon,ini}=500\,\Msun$ and the full sample ranging from $m_{\rm baryon,ini}=30-7100\,\Msun$.
Thus, the isolated galaxies are generally better-resolved by 1-2 orders of magnitude, so they should not suffer (as much) from numerical over-quenching.

Figure ~\ref{fig:qf_isolated} shows the quiescent fraction of these high-resolution isolated galaxies compared to lower resolution satellites and centrals in the MW simulations.
Centrals are galaxies that are $1-2\,\mpc$ from the nearest MW-mass host.
We further require that centrals have never been within $R_{\rm 200m}$ of a more massive halo, because \citet{Jahn2021} showed that LMC-mass galaxies in FIRE-2 can environmentally quench their satellites, and we want to compare the high-resolution isolated galaxies to a sample of galaxies from the MW simulations that have not been environmentally quenched.
We note that one host (m12i) does not have any galaxies in our stellar mass range ($\mstarlowlim$) within $1-2\,\mpc$.

The satellite galaxies are more quiescent than centrals and isolated galaxies, as expected.
The comparison between high-resolution isolated galaxies and lower resolution centrals is more important, because it tests possible numerical effects of over-quenching in the MW-mass simulations in the absence of (strong) environmental effects.
At $\Mstar>10^6\,\Msun$, the quiescent fraction of isolated galaxies is consistent with that of centrals at the 1-sigma level, indicating that the ISM and star formation of these centrals are reasonably well-resolved.
However, all three isolated galaxies with $\Mstar=10^{5-6}\,\Msun$ (m$_{\rm baryon,ini}=30-500\,\Msun$) are star-forming, whereas the centrals in our MW-mass simulations are typically quiescent.
The isolated sample also contains a single galaxy with $\Mstar<10^5\,\Msun$, which is quiescent.

The number of high-resolution isolated galaxies with $\Mstar<10^6\,\Msun$ is too small to draw a decisive conclusion on numerical issues, but the difference in quiescent fraction at $\Mstar=10^{5-6}\,\Msun$ between high-resolution isolated galaxies and lower-resolution centrals suggests that there is numerical over-quenching in the MW simulations at this mass scale.
Numerical over-quenching could be from the unphysical total ejection of gas in such low-mass galaxies, where just a few gas cells may constitute the ISM and receive all of the energy and momentum from a supernova event, causing them to be permanently removed from the galaxy.
Overall, galaxies in the MW simulations at $\Mstar\gtrsim10^6\,\Msun$ are likely well-resolved, but galaxies at $\Mstar<10^6\,\Msun$ are probably suffering from numerical over-quenching.
See \citet{Hopkins2018} for a detailed investigation of numerical effects in FIRE-2 galaxies.

\subsubsection{Comparisons to observations}\label{sec:qf_obs}

Here, we compare the quiescent fraction of simulated satellites to the LG and the SAGA survey.
We note that in this section we show host-to-host average and scatter rather than statistical quiescent fraction and uncertainty for the simulations, and we use different definitions of quiescence that depend on the observations we compare to.
We emphasize that we selected the simulated MW host halo masses to lie within reasonable observational estimates of the MW's halo mass ($\Mtwohm\approx1-2\times10^{12}\,\Msun$).
The SAGA hosts were chosen to be MW analogs based in part on $-23>M_K>-24.6$ ($\Mstar\approx10^{10-11}\,\Msun$), and consistent with the stellar masses of our simulated hosts and the MW.
SAGA systems should provide a statistical sample of observed MW analog satellites to contextualize the simulation results.

In Figure~\ref{fig:qf_obs} (left), we compare the simulations to LG satellites.
We include observational results for the LG, where quiescence is defined based on the absence of detectable HI, updated from \citet{Wetzel2015b} using data from \citet{Putman2021}.
Error bars on the observational LG points are statistical estimates of uncertainty that are described in Section~\ref{sec:q_def}..
We select simulated satellites ($\dlim$) based on their 3D distance from the nearest MW-mass host, and we define quiescence here as $\MHI<10^6\,\Msun$ to fairly compare to the LG observations.
We generate the simulation results by finding the fraction of quiescent satellites in 1 dex bins of stellar mass over $\Mstar=10^{5-10}\,\Msun$ for each host, and we present the host-to-host mean and scatter (68 and 100 per cent).
The scatter is quite large and many hosts have quiescent fractions at the upper and lower extremes of the distribution, which is why the 68 and 100 per cent limits of the distribution coincide at $\Mstar=10^{7-9}\,\Msun$.

Notably, given the large host-to-host variations, the simulations are broadly consistent at the 1-sigma level with the LG across the full range of satellite stellar masses. 
At lower masses ($\Mstar<10^7\,\Msun$), both the simulations and the LG have almost completely quiescent satellite populations.
At $\Mstar\approx10^{7-9}\,\Msun$, the simulation average is more star-forming than the LG, but the simulation average and LG data converge again in the highest mass bin ($\Mstar=10^{9-10}\,\Msun$) where all satellites are star-forming.

In Figure~\ref{fig:qf_obs} (right), we compare our simulations to the SAGA survey.
We take results for the SAGA survey from \citet{Mao2021}.
SAGA's definition of quiescence is based on the absence of significant H$\alpha$ emission, which probes star formation on short timescales, so here we use only our time criterion for quiescence: no star formation in a galaxy within the last 200 Myr.
The inner (darker) error bars on the SAGA data are shot noise, and the outer (lighter) error bars are an estimate of the impact of maximal incompleteness, by making the assumptions that the photometry of satellite candidates is complete to spectroscopic limits and that all spectroscopic non-detections are quiescent.

We select simulated galaxies such that they lie within 300 projected kpc from the host.
We allow foreground and background galaxies within $\pm2$ Mpc of the host to scatter within the mock survey area if their line-of-sight velocities are within $\pm275\,\rm{km/s}$ of the host's velocity, which is the same velocity cut used by SAGA \citep{Mao2021}.
We perform this mock observation along 1000 lines of sight per host, so the scatter in this panel is representative of both host-to-host and line-of-sight variations.
Our projected selection allows some non-satellites to be counted in the satellite sample, which are more likely to be star-forming than satellites of similar mass (see Figure~\ref{fig:qf_isolated}).
This shifts the mean simulation quiescent fraction down about 0.1 for $\Mstar=10^{8-9}\,\Msun$ and moves the lower bound of the distribution for $\Mstar=10^{6-7}\,\Msun$ down by 0.3. 
However, we also note that the mean simulation quiescent fraction actually increases by about 0.1 for $\Mstar\approx10^{7-8}\,\Msun$, away from the SAGA data.
The lowest-mass bin of SAGA is more consistent with the lower bound of the simulation distribution in this panel compared to the left panel, and given the large estimated observational uncertainty.
Thus, one partial explanation for the difference between the measured quiescent fractions in the LG and the SAGA survey could be the inclusion of non-satellites in the SAGA sample.

Overall, our results reproduce the average trend in the LG quiescent fraction; lower-mass satellites are mostly quiescent, roughly half of intermediate-mass satellites are quiescent, and higher-mass satellites are mostly star-forming.
Almost all simulated satellites at $\Mstar<10^7\,\Msun$ are quiescent, so the turnover in the SAGA survey's quiescent fraction at $\Mstar\lesssim10^{7.5}\,\Msun$ is conspicuous in comparison.
However, \textit{some} simulated hosts have overwhelmingly star-forming satellites at $\Mstar\approx10^{7-9}\,\Msun$, similar to the SAGA survey, and we note that the host-to-host average quiescent fraction in the simulations lies roughly between the LG and the SAGA survey (both panels of Figure~\ref{fig:qf_obs}).
Combined with the fact that the host-to-host scatter is large enough to permit almost any quiescent fraction at intermediate satellite stellar masses, our results may imply that differences in MW-mass host environments in the observed samples could lead to significantly different quiescent fractions.
We explore some possible physical differences in host environments that could lead to disparate quiescent fractions of satellites in the next section.

\begin{figure*}
    \begin{multicols}{2}
	\includegraphics[width=0.46\textwidth]{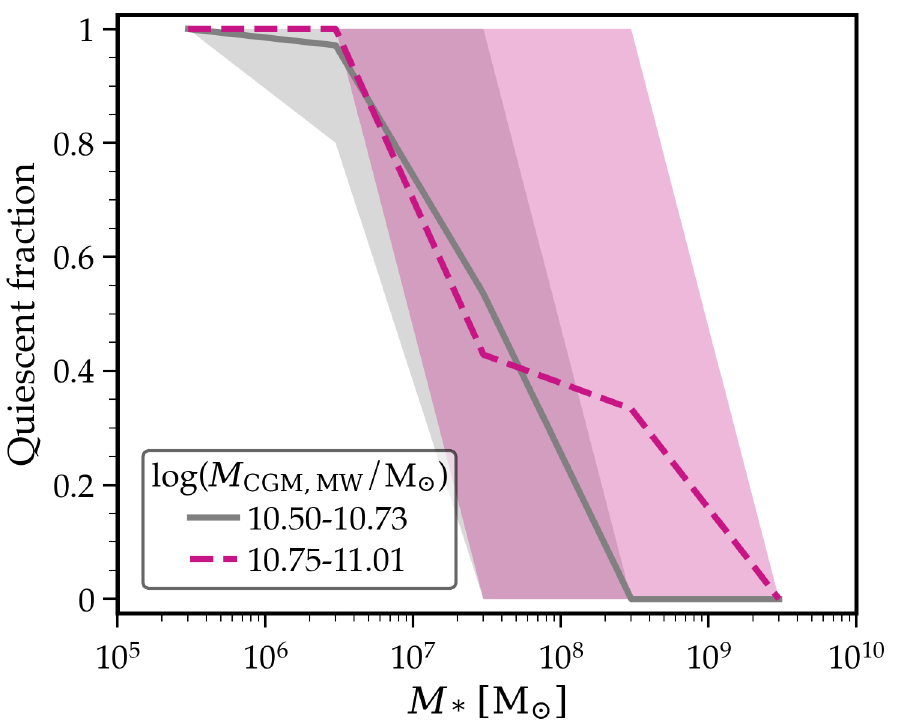}\par
	\includegraphics[width=0.46\textwidth]{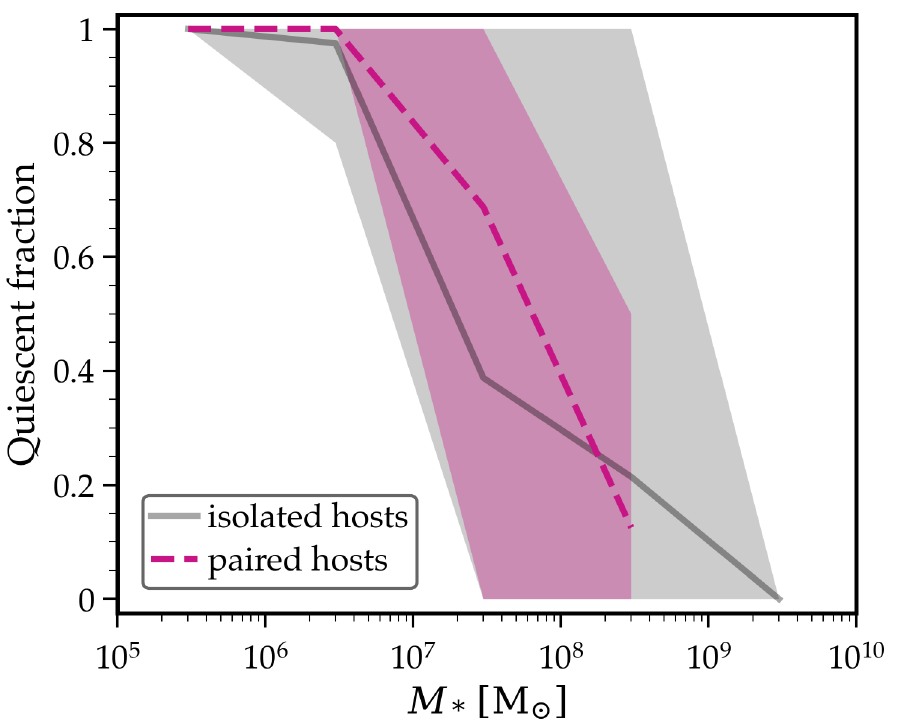}\par
	\end{multicols}
	\vspace{-6 mm}
    \caption{
    Trends in the quiescent fraction of satellites with properties of the MW-mass host environment. 
    The lines and shaded regions are the host-to-host means and 100 per cent scatters for each subsample.
    \textit{Left:} We split our sample of 14 hosts by the mass of their circumgalactic medium (CGM).
    The hosts with more massive CGM tend to have a larger quiescent fraction of massive satellites ($\Mstar=10^{8-9}\,\Msun$) that scatters up to fully quiescent.
    Whereas, the hosts with less massive CGM have a fully star-forming population of satellites at $\Mstar=10^{8-9}\,\Msun$.
    We find a similar, but weaker, trend with total host mass (not shown).
    Overall, massive satellites of hosts with higher CGM or total mass are slightly more quiescent on average.
    \textit{Right:} We divide our sample of hosts by paired/LG-like (6 hosts) or isolated (8 hosts) environment.
    The scatter for paired hosts is slightly narrower and the mean is higher ($+0.3$) at $\Mstar=10^{7-8}\,\Msun$, but given the large scatter in each subsample, the difference between quiescent fractions around paired and isolated hosts is not significant.
    }
    \label{fig:QF_hostenv}
\end{figure*}

\subsubsection{MW-mass host environment}\label{sec:qf_host_env}

We now explore some possible astrophysical explanations for the large host-to-host scatter in the quiescent fraction of satellites within our simulations.
We investigate the impact of the MW/M31-mass host environment by separating our sample of 14 hosts by CGM gas mass, total mass ($\Mtwohm$), and isolated versus paired host environment.
Given that the SAGA survey targets satellites around isolated MW-mass hosts rather than paired MW-mass hosts, this could contribute to the different quiescent fractions in the LG and SAGA.
We present an overview of trends with host environment in this section and we show the quiescent fraction of each individual host’s satellite system in Appendix~\ref{sec:appendix_q_def}.

In Figure~\ref{fig:QF_hostenv} (left), we show the host-to-host average quiescent fraction and 100 per cent scatter for the low and high CGM mass groups.
We measure CGM mass by summing all gas mass within $30-300\,\kpc$ of the MW-mass hosts, excluding gas that is assigned to the satellites and within the inner halo where we do not typically find surviving satellites \citep{Samuel2020}.
We calculate a quiescent fraction for each host and then divide our sample in half: seven low CGM mass hosts ($3.1\times10^{10}\, \Msun<\rm{M_{CGM}}\leq5.5\times10^{10}\,\Msun$) and seven high CGM mass hosts ($5.5\times10^{10}\,\Msun<\rm{M_{CGM}}\lesssim1.0\times10^{11}\,\Msun$).
The difference in the mean CGM masses of each sample is about 60 per cent relative to the low CGM mass average, and the most massive CGM is about 3.2 times more massive than the least massive.
On average, hosts with higher CGM mass have massive satellites ($\Mstar=10^{8-9}\,\Msun$) that are nearly 40 per cent quiescent, though the scatter ranges from $0-100$ per cent quiescent in this bin.
In contrast, hosts with lower CGM mass have a fully star-forming satellite population in the same satellite mass range.
This suggests that a higher CGM mass quenches massive satellites more efficiently (probably because of enhanced ram pressure), but lower-mass satellites ($\Mstar\lesssim10^{7}\,\Msun$) are easily quenched by all CGM masses in our host sample.
However, we note that our result here is somewhat marginal because not all hosts have satellites at $\Mstar=10^{8-9}\,\Msun$ and we find only a weak trend in the time between infall and quenching with CGM mass.

We also examine whether the total mass ($\Mtwohm$, which includes stars, gas, and dark matter) is connected to satellite quenching.
We divide our host sample in half, which yields seven lower-mass hosts ($9.2\times10^{11}\,\Msun<\Mtwohm\leq1.3\times10^{12}\,\Msun$) and seven higher-mass hosts ($1.3\times10^{12}\,\Msun<\Mtwohm<2.1\times10^{12}\,\Msun$).
The difference in the mean total masses of each sample is about 40 per cent relative to the low total mass average, and the most massive host is about 2.2 times more massive than the least massive.
We find a similar trend compared to separating hosts by CGM mass, whereby the fraction of quiescent satellites with $\Mstar=10^{8-9}\,\Msun$ is about 0.2 higher around massive hosts than around less massive hosts.
The trend is less significant in this case, because the host-to-host scatter ranges from $0-100$ per cent for both host samples in this bin.

In Figure~\ref{fig:QF_hostenv} (right), we separate hosts by whether they are a member of a LG-like paired system from the ELVIS on FIRE suite (6 hosts) or an isolated host from the Latte suite (8 hosts).
We find that the paired hosts have a higher mean quiescent fraction by 0.3 in the $\Mstar=10^{7-8}\,\Msun$ mass bin, and a slightly lower quiescent fraction in the $\Mstar=10^{8-9}\,\Msun$ mass bin.
However, the scatter in paired hosts is completely encompassed by the larger scatter in isolated hosts, so this difference is likely not significant.
The paired hosts also lack more massive satellites to compare against in our highest mass bin.
In light of our results in the left panel of Figure~\ref{fig:QF_hostenv}, we check for offsets in the average CGM and total host masses of the paired and isolated host samples, but we do not find significant differences.
In Section~\ref{sec:quench_times}, we also check the infall and quenching times of satellites amongst the different host environments, and the only difference we find is slightly earlier infall times ($\approx1\,\gyr$) for satellites of the paired hosts.
See Section~\ref{sec:quench_times} for more details.

Our results on host mass effects generally agree with previous studies of simulations and observations that leverage larger samples of MW-mass hosts.
\citet{Font2022} found lower quiescent fractions of satellites ($\Mstar\approx10^{6.5-9.5}\,\Msun$) around less massive hosts in the ARTEMIS simulations (45 hosts with $8 \times 10^{11} < \Mtwohm/\Msun < 2 \times 10^{12}$). 
\citet{Carlsten2022} also found a similar trend in observations where the overall quiescent fraction of satellites ($M_V<-9$ mag, $\Mstar\gtrsim5\times10^5\,\Msun$) increases with the luminosity of observed MW-like hosts (30 host galaxies with $M_{K_s} < -22.1$ mag) in the Local Volume ($D<12\,\mpc$).
Taken together, these results could indicate that host-mass-dependent quenching could be universal, and based on our results, the host CGM mass (which should correlate with total mass) may drive this trend, likely because of enhanced ram pressure at higher CGM masses.
If SAGA is somehow preferentially selecting hosts with less massive CGM or total mass, then we might expect to see a smaller quiescent fraction of satellites at $\Mstar\gtrsim10^8\,\Msun$.
However, further analysis is needed to determine the specific cause of enhanced quenching around more massive hosts.

\begin{figure*}
    \begin{multicols}{2}
    \includegraphics[width=0.46\textwidth]{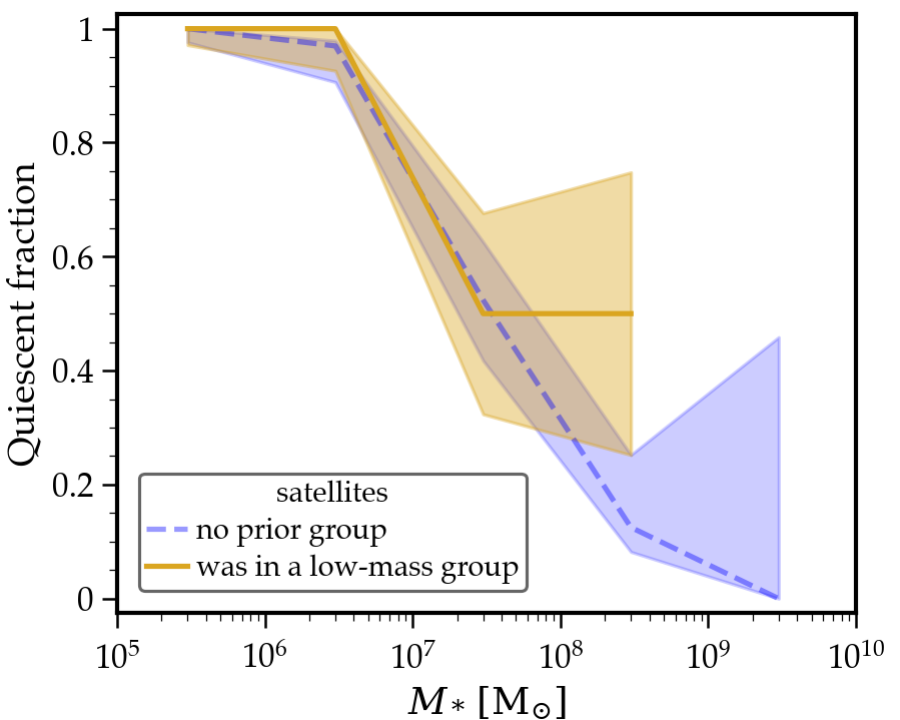}\par
    \includegraphics[width=0.46\textwidth]{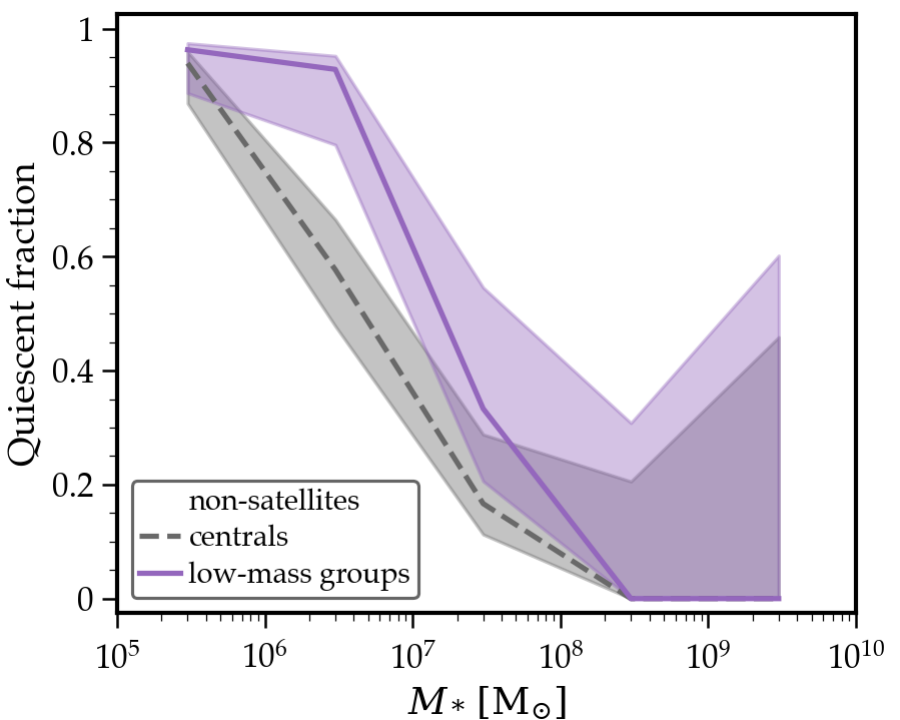}\par
    \end{multicols}
    \vspace{-6 mm}
    \caption{Effects of environmental quenching in low-mass groups on the quiescent fraction of satellite ($\dlim$) and non-satellite ($1-2\,\mpc$ from the nearest MW-mass galaxy) galaxies.
    Lines stack all galaxies across the different simulations together.
    \textit{Left:} Satellites at $\Mstar=10^{8-9}\,\Msun$ that were previously in a group are more likely to be quiescent than satellites of similar mass that did not have prior group associations.
    \textit{Right:} Non-satellites at $\Mstar=10^{6-8}\,\Msun$ that are part of a low-mass group are more likely to be quiescent than centrals, which have never been within $\rtwohm$ of a more massive halo.
    Galaxies at $\Mstar<10^8\,\Msun$ can be environmentally quenched within low-mass groups, but this effect is obscured in the highly quiescent population of satellite galaxies around MW-mass hosts.
    }
    \label{fig:GP_qf}
\end{figure*}

\begin{figure}
    \centering
    \includegraphics[width=\columnwidth]{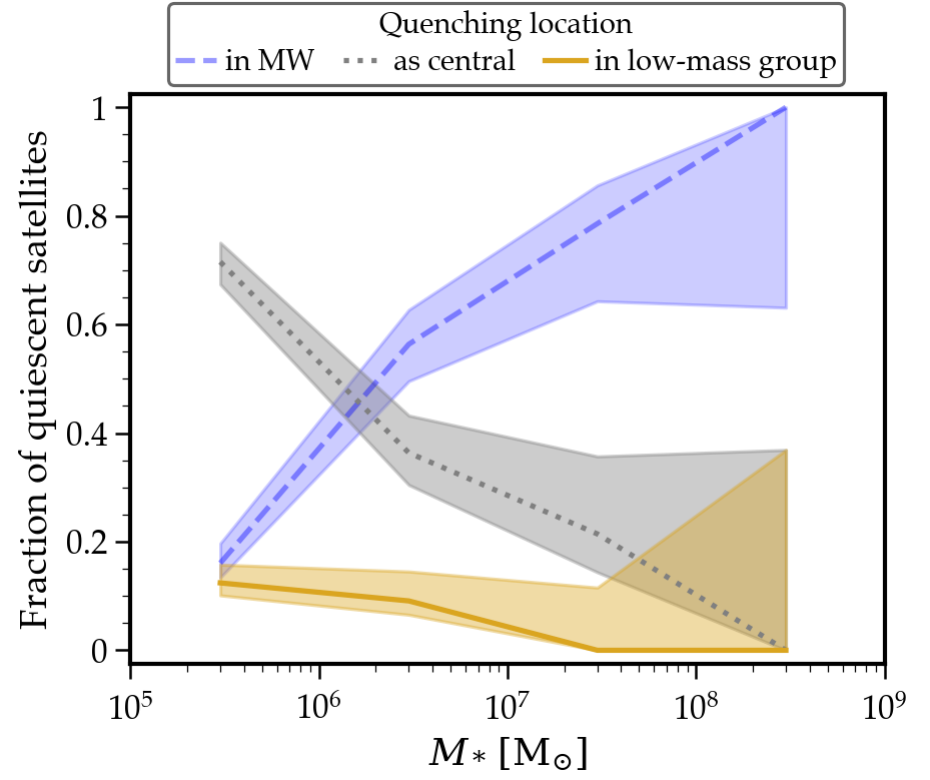}
    \vspace{-6 mm}
    \caption{Where satellite galaxies quench.
    We show the fraction of \textit{quiescent} satellites that quenched inside a MW-mass halo (blue), as a central before falling into a MW-mass halo (grey), or within a low-mass group before falling into a MW-mass halo (yellow).
    Satellites at $\Mstar>10^6\,\Msun$ mostly quenched within the MW-mass halo.
    About 10 per cent of satellites with $\Mstar<10^7\,\Msun$ quenched within a low-mass group before entering the MW halo.
    Satellites at $\Mstar=10^{5-6}\,\Msun$ in our simulations typically quenched as centrals, likely from a combination of cosmic reionization, stellar feedback, and resolution effects.
    Quenching within a MW-mass halo dominates for well-resolved satellites ($\Mstar>10^6\,\Msun$), but preprocessing in low-mass groups also contributes non-negligibly to the quiescent fraction of satellites at $\Mstar<10^7\,\Msun$.
    }
    \label{fig:quench_location}
\end{figure}

\subsubsection{Group preprocessing}\label{sec:GP}

Ram pressure within a MW-mass halo should be sufficient to quench most satellites with $\Mstar\lesssim10^8\,\Msun$, as long as the host CGM is sufficiently clumpy \citep{Fillingham2016}.
However, environmental preprocessing in a lower-mass group, prior to falling into a MW-mass halo, may also contribute to the quiescent population of MW satellites \citep{Li2008,Wetzel2013,Wetzel2015a,Deason2015}.
In light of the environmental quenching of satellites around isolated LMC-mass galaxies from the FIRE-2 simulations \citep{Jahn2021}, we search for signs of group preprocessing in low-mass groups prior to infall into the MW-mass haloes in our simulations.

We identify prior group associations of $z=0$ satellites using the merger tree of each simulation.
The merger trees record when a halo passes within a more massive halo's radius ($R_{\rm 200m}$).
If one of the satellites was within a halo more massive than itself prior to MW infall, we count it as being in a low-mass group.
We find that 93 ($\approx40$ per cent) of our 240 satellites had such a prior group association.
The fraction of satellites that were in low-mass groups is the same around the paired and isolated MW-mass hosts.
These low-mass groups have a variety of histories; some of them stay bound within the MW halo and others separate before MW infall.

Figure~\ref{fig:GP_qf} (left) shows the quiescent fraction of satellites, separating satellites by whether or not they were in a low-mass group.
At $\Mstar=10^{8-9}\,\Msun$, satellites with prior hosts are more quiescent than those without prior hosts, though there are few satellites in this mass bin.
Though the quiescent fraction of satellites with and without prior group associations is similar at $\Mstar<10^8\,\Msun$, it does not indicate where or when they quenched.
The left panel of Figure~\ref{fig:GP_qf} includes one splashback/renegade galaxy in the `no prior group' sample that is a satellite of one paired MW-mass host and has orbited within the halo of the other MW-mass host in that simulation, but it has no prior low-mass group association.
In Figure~\ref{fig:GP_qf} (right), we also show that non-satellite galaxies ($1-2\,\mpc$ from the nearest MW-mass host) that have been in a low-mass group are more quiescent than centrals at $\Mstar=10^{6-8}\,\Msun$.
The right panel of Figure~\ref{fig:GP_qf} does not include any splashback galaxies of MW-mass hosts, as the centrals in the right panel have never been within the halo of \textit{any} more massive host and the non-satellites in low-mass groups have been selected to exclude hosts with halo masses $\Mtwohm<10^{12}\,\Msun$.
However, the solid purple line may include splashback galaxies of low-mass hosts, as we do not require the low-mass group associations to persist to $z=0$.
Our results for non-satellite galaxies signal that group preprocessing could significantly contribute to the quenching of MW satellites.

In Figure~\ref{fig:quench_location}, we investigate the impact of group preprocessing versus MW environmental quenching further by examining where quiescent satellites had their last episode of star formation.
We show the fraction of quiescent satellites that quenched within the MW halo radius, outside of the MW halo as a central, and in low-mass groups (within a low-mass group host's halo radius).
Most quiescent satellites with $\Mstar>10^6\,\Msun$ quenched within the MW host halo, especially higher-mass satellites.
Although 40 per cent of all satellites were in a low-mass group, only about 10 per cent of satellites with $\Mstar<10^7\,\Msun$ quenched within a low-mass group.
However, that does not necessarily mean the others are not preprocessed by their small groups (see Case III in Section~\ref{sec:case_studies}, which quenches as a central but is clearly preprocessed on a low-mass group).
Most of the satellites with $\Mstar<10^6\,\Msun$ quenched as centrals, likely from a combination of reionization, stellar feedback, and resolution effects.
Group preprocessing does not dominate whether or where a MW satellite quenches, but it can contribute to the fraction of quiescent satellites (about 10 per cent at $\Mstar<10^7\,\Msun$) and quicken the process of quenching upon infall into the MW halo, which we explore in the next section.

We do not find any significant dependence of the quiescent fraction of satellites on the dark matter halo mass of the prior host, but our sample of galaxies with prior hosts is limited at $\Mstar>10^7\,\Msun$, where we have both quiescent and star-forming galaxies, so we are unable to draw a strong conclusion.
However, the quiescent fraction of non-satellites that have been in groups rises slowly with the dark matter halo mass of the group host for halo masses $\approx10^{7-11}\,\Msun$.
This could imply a similar trend with host mass as that found in Section~\ref{sec:qf_host_env}, but a more detailed analysis of low-mass group hosts and their CGM is necessary to confirm this.

\subsection{Quenching timescales}\label{sec:quench_times}

To investigate the coincidence in time between quenching and specific environmental events, we define a metric for each satellite that we refer to as the quenching delay time: $t_{\rm quench}-t_{\rm env}$, which measures the time it takes for a satellite to quench relative to an event like infall into a host halo or pericentre passage.
$t_{\rm quench}$ is the time of last star formation in a galaxy, which is an `archaeologically' defined quantity in the simulation output, that is, each star particle has an intrinsic age corresponding to its formation time that may have occurred between saved snapshots.
A quenching delay time near zero indicates that a satellite quenched close to the particular event and a positive (negative) value indicates that the satellite quenched after (before) the event.

\subsubsection{Infall}

Analytical and numerical estimates of quenching timescales have shown that satellite galaxies of intermediate masses ($\Mstar\approx10^{6-8}\,\Msun$) are often quenched by the host environment within about 2 Gyr of infall into a MW-mass halo, and that quenching takes longer in more massive satellites, with few quenched satellites at $\Mstar\gtrsim10^9\,\Msun$ \citep{Wetzel2015b,Fillingham2016,RodriguezWimberly2019}.

\begin{figure}
    \centering
    \includegraphics[width=\columnwidth]{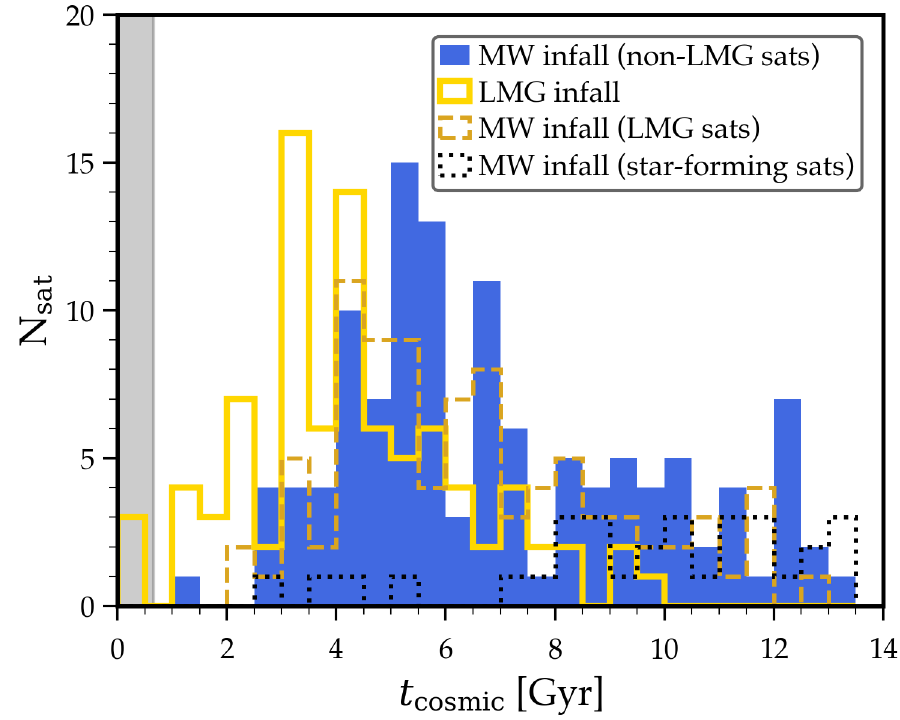}
    \vspace{-6 mm}
    \caption{Infall times for all satellite galaxies in the simulations.
    We show quiescent satellites as colored histograms and star-forming satellites as a dotted black histogram.
    For satellites that were in low-mass groups (LMGs), we show their infall times into the low-mass host (solid yellow line) and their infall times into the MW-mass host (dashed dark yellow line).
    The grey shaded region shows the era of HI reionization in the simulations that ends at $z\approx8$.
    A few satellites that were in low-mass groups have infall times within the HI reionization era, but the majority of them have infall times a few to several Gyr after reionization ends.
    Star-forming satellites fell into the MW-mass host later on average compared to quiescent satellites.
    }
    \label{fig:infall_times}
\end{figure}

Figure~\ref{fig:infall_times} shows the infall times of satellite galaxies in the simulations.
We define infall as when a galaxy first crosses the radius of a more massive host halo: $\dhost\leq R_{\rm 200m,\,host}$.
For satellites that were in low-mass groups,
we show their infall times into both the low-mass host and the MW-mass host haloes.
Some satellites that were in low-mass groups have infall times during or near the HI reionization era that ends at $z\approx8$, which could potentially contribute to their quenching apart from group preprocessing, but the vast majority of satellites that were in low-mass groups have infall times a few to several Gyr after reionization ends.
We have also verified that even when we exclude all satellites that were in low-mass groups that quench before $t=4\,\gyr$ our results from Section~\ref{sec:GP} are qualitatively unchanged.

\begin{figure*}
    \begin{multicols}{2}
	\includegraphics[width=0.47\textwidth]{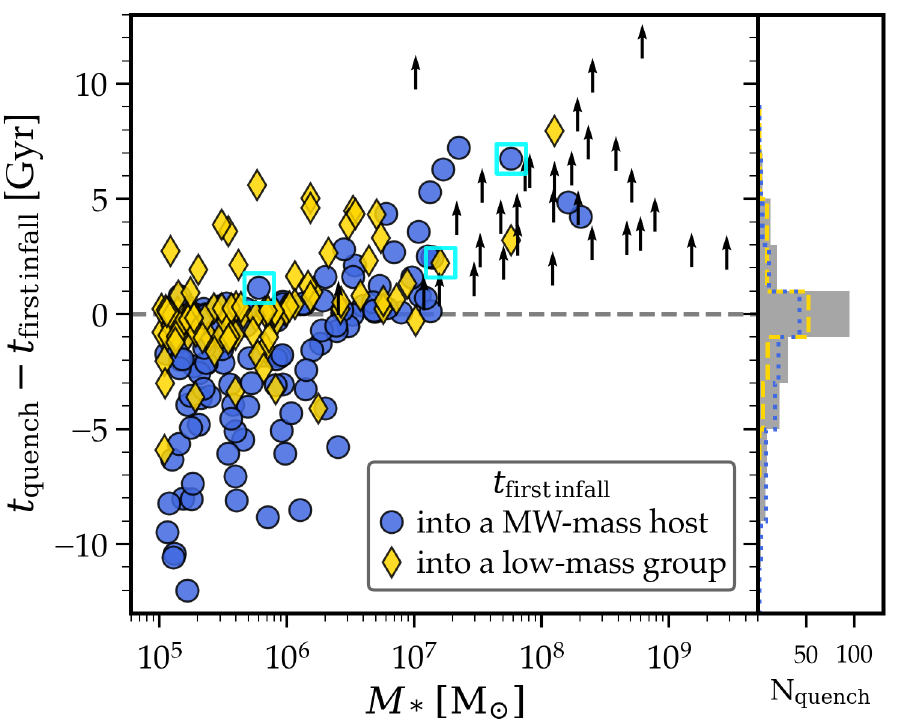}\par
	\includegraphics[width=0.47\textwidth]{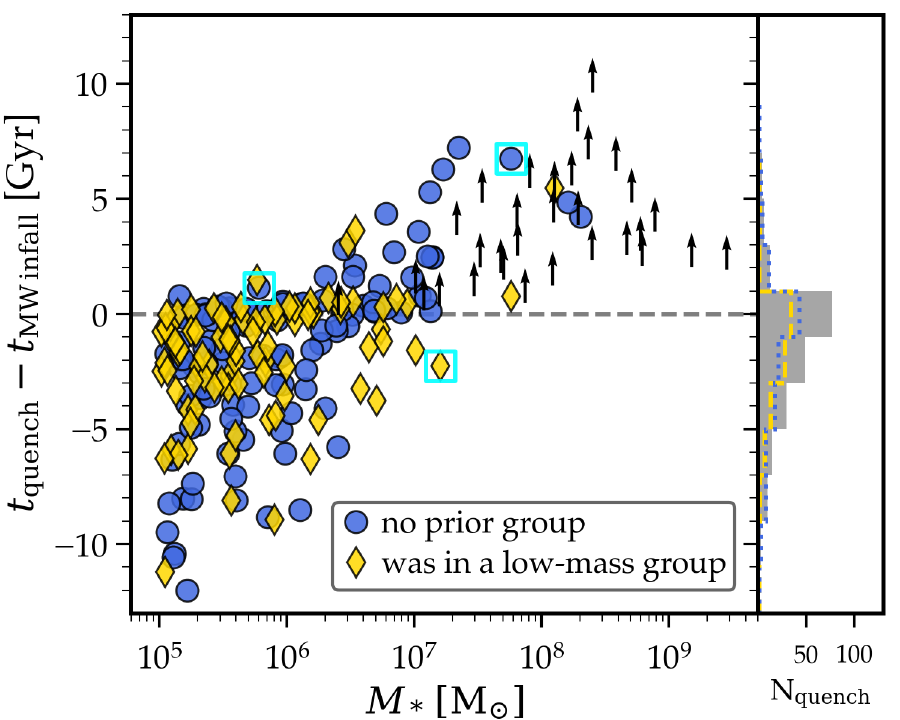}\par
	\end{multicols}
	\vspace{-6 mm}
    \caption{When satellites quench relative to different infall events.
    We show quiescent satellites as filled symbols and star-forming satellites as arrows to indicate a lower limit on their quenching delay times.
    Yellow diamonds are satellites that were in a low-mass group before entering the MW halo, and blue circles had no prior group association.
    In general, the typical quenching delay time increases with satellite stellar mass, similar to previous studies.
    \textit{Left:} Quenching delay time versus stellar mass relative to \textit{first} infall into \textit{any} host for all satellites at $z=0$.
    Most quiescent satellites have a quenching delay time near zero, demonstrating that they rapidly quenched near first infall.
    All satellites at $\Mstar\gtrsim10^{6.5}\,\Msun$ quenched during or after falling into a host halo, which could imply an upper mass limit on isolated quenching in low-mass galaxies.
    \textit{Right:} Same as left, but quenching delay time is relative to infall into \textit{a MW-mass host}.
    Compared to the left panel, the satellites that were in low-mass groups do not cluster as strongly near zero and many of them move to lower values on the y-axis.
    This demonstrates that some cases of negative quenching delay time with respect to MW infall are actually because of group preprocessing.
    We show the evolution of the three satellites highlighted with cyan squares in Section~\ref{sec:case_studies}.
    }
    \label{fig:QDT}
\end{figure*}

Figure~\ref{fig:QDT} (left) shows quenching delay time relative to \textit{first infall into any host halo} (including the MW-mass hosts and low-mass group hosts similar to the LMC and SMC) versus stellar mass for all 240 satellite galaxies in our simulations.
More massive galaxies tend to quench after first infall and less massive galaxies quench before or during first infall.
All satellites with $\Mstar\gtrsim10^{8.5}\,\Msun$ are star-forming.
These trends are generally consistent with the semi-empirical results in \citet{Wetzel2015b} and \citet{Fillingham2016} as well as results from baryonic cosmological simulations such as \citet{Akins2021}.

There is a noticeable build-up of low-mass galaxies ($\Mstar\lesssim10^7\,\Msun$) along the line of zero quenching delay time, showing that many satellites quenched during first infall and that rapid environmental quenching from infall is strongest in this mass range.
We find that 45 (60) per cent of all quiescent satellites cease star formation within $\pm1\,\gyr$ ($\pm2\,\gyr$) of infall into any host.
Quenching delay times for low-mass satellites extend to large negative values (and early cosmic times, see Appendix~\ref{sec:appendix_QDT}) likely because these galaxies are quenched by a combination of reionization, stellar feedback, and resolution effects.
It is also notable that almost no satellites above $\Mstar\approx10^{6.5}\,\Msun$ quench before infall into any host.
This may imply that galaxies above this mass threshold do not self-quench in general, or as a feature of these particular simulations.

Some satellites that were in low-mass groups also tend to have large positive quenching delay times.
This spread in quenching delay times could be due to properties of the low-mass group or aspects of the interaction between the satellite and the low-mass group.
However, we find no significant trends in the quenching delay times of these satellites with low-mass group host halo mass, closest approach to low-mass group host, and maximum relative velocity between the satellite and the low-mass group host. 
The lack of any clear trend is somewhat surprising, but quantifying the effects of an interaction with a low-mass group likely requires a dedicated analysis of ram pressure and tidal disruption.

In Figure~\ref{fig:QDT} (right), we show the quenching delay time with respect to infall into \textit{only a MW-mass halo}.
Intermediate-mass ($\Mstar=10^{7-8}\,\Msun$) quiescent satellites quenched within $\approx2.5\,\gyr$ of MW infall on average.
Few of the massive satellites ($\Mstar\gtrsim10^8\,\Msun$) have quenched by $z=0$, and those that did quenched $\approx5\,\gyr$ after MW infall on average.
Of the 70 satellites that quenched after MW infall, 71 per cent (53 per cent) quench within 1 Gyr (500 Myr) of infall.
Notably, all satellites with $\Mstar\lesssim10^{6.5}\,\Msun$ that quenched after MW infall did so within 2 Gyr of infall, indicating rapid environmental effects.
Our quenching delay time for massive satellites is faster than the $\approx8\,\gyr$ found by \citet{Fillingham2015}, but consistent with the low end of $\approx5-10\,\gyr$ from \citet{Wetzel2015a}.
This could point to stronger or more effective environmental quenching of massive satellites in our simulations than implied by those earlier works.

Comparing the histograms on the left and right panels of Figure~\ref{fig:QDT}, satellites that were in low-mass groups are more clustered near zero in the left panel (first infall) than the right panel (MW infall).
The fraction of satellites that quench within $\pm1\,\gyr$ of MW infall is just 37 per cent compared to 45 per cent that quench within $\pm1\,\gyr$ of first infall into any host.
The quenching delay times of several satellites that were in low-mass groups also shift from negative when measured with respect to MW infall to positive when measured with respect to first infall into any host.
This indicates that some satellites were preprocessed or began quenching in a low-mass group before MW infall, and signals the importance of considering group preprocessing, to avoid a mis-interpretation of a (potentially unphysical) negative quenching delay time with respect to MW infall.
We note that when we exclude all satellites that were in low-mass groups that quench before $t=2\,\gyr$ (to avoid potential lingering contamination from reionization that formally ends after 1 Gyr) these trends persist.
However, if we exclude any low-mass group satellites that quenched before $t=4\,\gyr$, then the trends are mostly washed out, in part due to a significant reduction in the number of satellites used for the comparison.
Our results are consistent with \citet{Jahn2021}, who found that satellites of isolated LMC-mass galaxies can be efficiently environmentally quenched on short timescales ($\lesssim2-4\,\gyr$) in the FIRE-2 simulations.

Furthermore, intermediate-mass satellites ($\Mstar=10^{7-8}\,\Msun$) that were in low-mass groups also have a lower median (68 per cent) quenching delay time with respect to MW infall of 0.8 Gyr (-1.8 to 1.2 Gyr), compared to satellites that were not in a low-mass group (median of 2.5 Gyr and 68 per cent range of 0.7 to 6.3 Gyr).
This likely reflects that environmental quenching in low-mass groups can significantly quicken the process of quenching within a MW-mass halo.

In light of the dependence of quenching on MW-mass host properties that we find in Section~\ref{sec:qf_host_env}, we check the infall and quenching times of satellites amongst the different host subsamples that we considered, because this could skew the quiescent fraction if some satellites have been within the host environment for a longer time.
Infall times are similar for hosts with high or low CGM and total masses (median infall times are offset by $\lesssim500\,\myr$).
The paired hosts have somewhat earlier infall times, with a median infall time of about 1.3 Gyr earlier and an early-infall tail reaching about 1 Gyr earlier than the isolated host distribution (Santistevan et al. in prep).
However, \citet{Wetzel2015a} analyzed a larger sample of dark matter-only versions of the isolated and paired MW-mass haloes and found no difference in their average subhalo infall times.
The fact that the paired host simulations have $\approx2$ times better mass resolution than the isolated host simulations could also mean that satellites of paired hosts are able to numerically survive for longer and hence the $z=0$ population would have earlier infall times on average.
Though we note that the distributions of cosmic times at satellite quenching are similar amongst paired and isolated hosts in our simulations, and the distributions of satellite quenching delay times (relative to infall into the MW-mass halo) are essentially the same for all our host subsamples.
While the difference in infall times between paired and isolated hosts is not large ($\approx1\,\gyr$ on average), it is comparable to the quenching delay times relative to MW infall for many lower-mass satellites ($\Mstar\lesssim10^7\,\Msun$).
Thus, we might expect (especially lower-mass) satellites of paired hosts to be more quiescent. 
Though we find a modest enhancement in the quiescent fraction of satellites around paired hosts at $\Mstar=10^{7-8}\,\Msun$, satellites below these masses are essentially uniformly quiescent so we conclude that in general there is no significant difference in the quiescent fraction around paired and isolated hosts.

\subsubsection{Pericentre passages}

We also explore quenching with respect to pericentre passages, because this is when satellites encounter the denser inner regions of the host CGM, and hence they should experience the most intense ram pressure.
We found pericentre passages by using a window search method for local minima over time in the distance of a satellite from the MW-mass host after it has crossed the host's radius for the first time and interpolating between snapshots with a cubic spline function (Santistevan et al. in prep).
We use a sufficiently large search window to eliminate spurious pericentre passages that may arise from snapshot cadence or numerical noise.
Though we do not find an especially strong correlation between quenching delay time with respect to first pericentre passage and satellite stellar mass compared to infall events (see Appendix~\ref{sec:appendix_QDT}), many satellites have experienced more than one pericentre passage.

\begin{figure}
    \centering
    \includegraphics[width=\columnwidth]{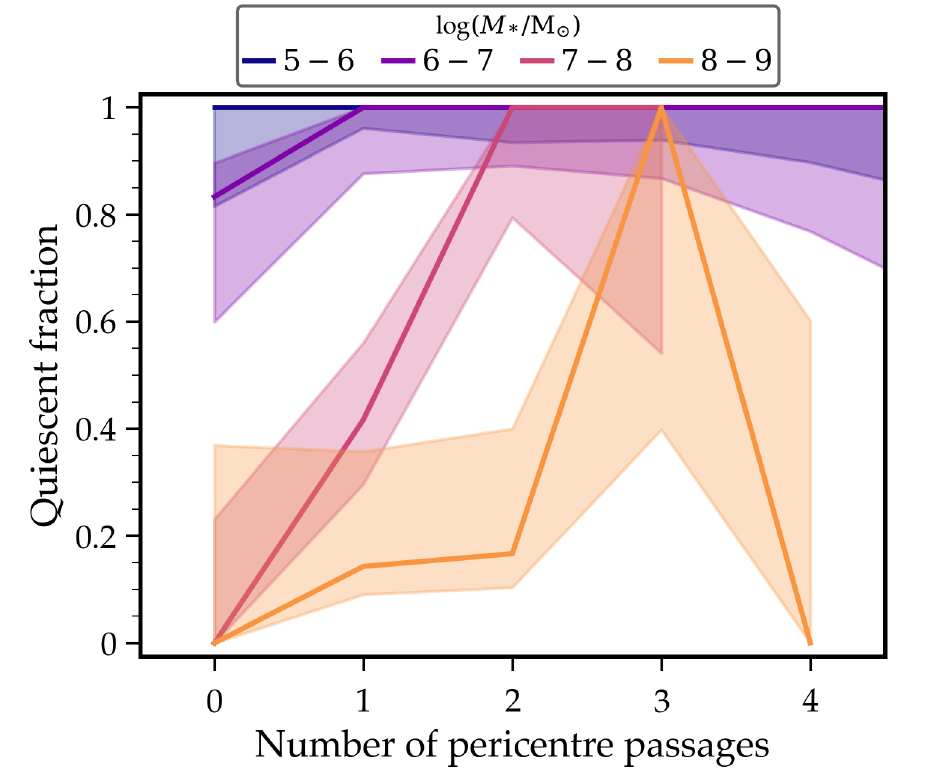}
    \vspace{-6 mm}
    \caption{The impact of pericentre passages on the quiescent fraction of satellites.
    We show the quiescent fraction of satellites at different stellar masses versus the total number of pericentre passages they have experienced.
    Satellites at $\Mstar=10^{5-7}\,\Msun$ are mostly quiescent regardless of the number of pericentre passages.
    The two satellites at $\Mstar>10^{9}\,\Msun$ have experienced one pericentre and are star-forming.
    One satellite at $\Mstar=10^{8-9}\,\Msun$ has experienced 4 pericentre passages and has not formed stars recently, but retains significant HI and is therefore star-forming according to our fiducial definition.
    In general, more massive satellites typically require multiple pericentre passages to quench.
    }
    \label{fig:QF_peris}
\end{figure}

In Figure~\ref{fig:QF_peris} we show the quiescent fraction of satellites versus the total number of pericentre passages they have experienced up to $z=0$.
About $80-100$ per cent of satellites at $\Mstar=10^{5-7}\,\Msun$ that have not yet experienced a pericentre passage are quiescent, and all satellites in this mass range that have experienced at least one pericenter passage are quiescent.
Satellites at $\Mstar=10^{7-9}\,\Msun$ show a more prominent dependence of quiescent fraction on the number of pericentre passages as the quiescent fraction rises steeply from 0 to 1 between 0 and 3 pericentre passages.
Out of the 17 satellites at $\Mstar=10^{7-9}\,\Msun$ that have experienced two or more pericentre passages, 11 (65 per cent) are quiescent.
Beyond a total of three pericentre passages, the satellite population at $\Mstar=10^{5-9}\,\Msun$ is completely quiescent (except for one satellite which still contains significant $\MHI$ but has not formed any stars in the last 200 Myr after four pericentre passages), indicating that three pericentre passages are likely sufficient to quench satellites at $\Mstar\leq10^{9}\,\Msun$.
The two highest-mass satellites in our sample ($\Mstar=10^{9-10}\,\Msun$, not shown in Figure~\ref{fig:QF_peris}) have each experienced a single pericentre passage and are still star-forming at $z=0$.
Several satellites at $\Mstar=10^{5-7}\,\Msun$ have actually experienced many pericentre passages (up to a maximum of 11), which we omit from the figure for clarity.

\subsubsection{Minimum quenching delay time}

\begin{figure}
	\includegraphics[width=0.47\textwidth]{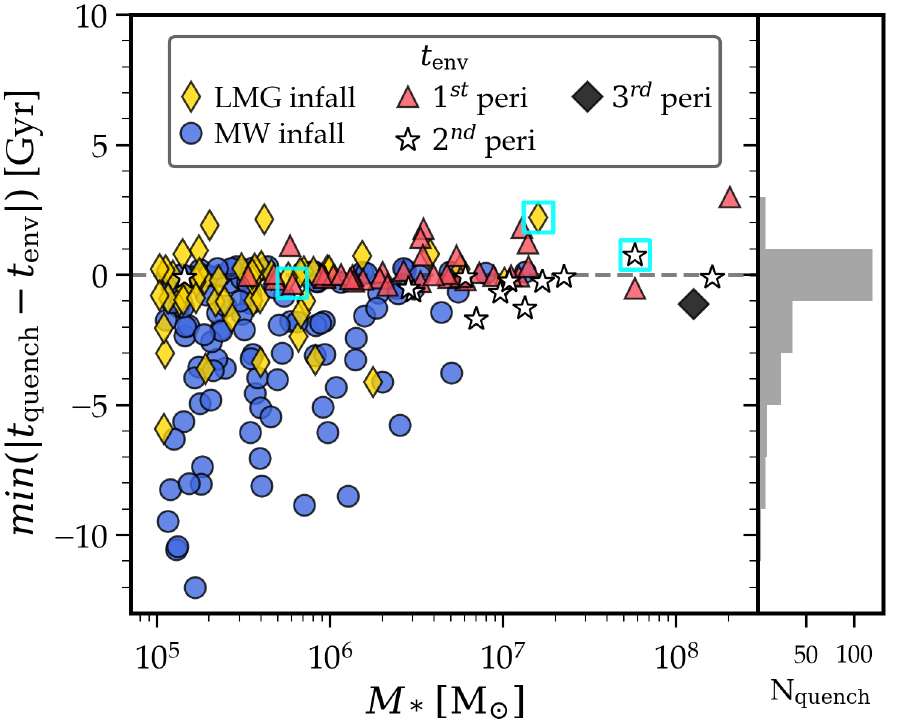}
	\vspace{-2 mm}
    \caption{Which environmental event occurs closest to quenching for every quiescent satellite.
    We show the minimum quenching delay time with respect to five different environmental events.
    LMG infall refers to infall into a low-mass group before infall into a MW-mass halo.
    Most quiescent satellites (73 per cent) quench within $\pm2\,\gyr$ of one of these events.
    There is a clear mass dependence in which event happens closest to quenching, where lower-mass satellites ($\Mstar\lesssim10^6\,\Msun$) typically quenched near infall and higher-mass satellites quenched after increasing numbers of pericentre passages.
    Satellites with large negative delay times typically quenched as centrals, which is closest to MW infall amongst the events we consider.
    We show the evolution of the three satellites highlighted with cyan squares in Section~\ref{sec:case_studies}.
    }
    \label{fig:QDT_min}
\end{figure}

Figure~\ref{fig:QDT_min} shows the minimum quenching delay time for all quiescent satellites, considering all infall and pericentre passage events.
We find that 57 (73) per cent of quiescent satellites quench within $\pm1\,\gyr$ ($\pm2\,\gyr$) of the infall or pericentre event that occurs closest to their quenching times.
The enhanced clustering of satellites near zero quenching delay time in this figure compared to the delay time for any \textit{single} type of event highlights the need to consider multiple types of events when determining the cause of satellite quenching.
Satellites with $\Mstar\lesssim10^{6}\,\Msun$ tend to quench close to infall, whereas more massive satellites quench closest to increasing numbers of pericentre passages.
We summarize the mass dependence of environmental quenching quantitatively in Table~\ref{table:QDT}.
The fact that many satellites quench closest to infall into a low-mass group (yellow diamonds) further demonstrates that effective environmental quenching of MW satellites can (begin to) occur in low-mass groups before MW infall.

\begin{table*}
\centering
\caption{The fraction of quiescent satellites that quenched closest to each type of environmental event that we consider. We highlight the two events corresponding to the highest fractions in each stellar mass bin. As satellite mass increases, the event closest to quenching moves from infall to increasing numbers of pericentre passages. Note that some low-mass satellites that quench closest to infall into a MW-mass halo actually quench as centrals before infall (see Section~\ref{sec:GP}).}
\begin{tabular}{lllllll}
\hline
 & & \multicolumn{5}{c}{Type of environmental event} \\
log$(\Mstar/\Msun)$ & $N_{\rm quench}$ & LMG infall & MW infall & $1^{st}$ peri & 2nd peri & 3rd peri \\
\hline
5--6 & 137 & \textbf{0.32} & \textbf{0.61} & 0.06          & 0.01          & 0 \\
6--7 & 55  & 0.07          & \textbf{0.44} & \textbf{0.40} & 0.09          & 0 \\
7--8 & 14  & 0.14          & 0.07          & \textbf{0.43} & \textbf{0.36} & 0 \\
8--9 & 3   & 0             & 0             & 0.33          & 0.33          & 0.33 \\
\hline
\end{tabular}
\label{table:QDT}
\end{table*}

\begin{table}
\centering
\caption{Properties of the three satellites that we explore as case studies. We give all masses in $\Msun$ and we specify them at $z=0$ except for $M_{\rm peak, halo}$, which is at an arbitrary time typically prior to first infall. None of these galaxies contain HI at $z=0$. The fifth column is the minimum quenching delay time (and corresponding closest event to quenching) amongst the different infall and pericentre passage events that we consider in Section~\ref{sec:quench_times}. LMG infall refers to infall into a low-mass group before infall into a MW-mass halo.}
\begin{tabular}{l|llll}
\hline
Case & $\Mstar$ & $\Mgas$ & $M_{\rm peak, halo}$ & QDT [Gyr] \\ \hline
I & $6 \times 10^5$ & $1 \times 10^4$ & $4 \times 10^9$ & -0.30 ($1^{st}$ peri) \\ 
II & $6 \times 10^7$ & 0 & $1 \times 10^{10}$ & +0.75 ($2^{nd}$ peri) \\ 
III & $2 \times 10^7$ & 0 & $9 \times 10^{9}$ & +2.20 (LMG infall) \\ \hline
\end{tabular}
\label{table:case_studies}
\end{table}

\subsection{Case studies}\label{sec:case_studies}

\begin{figure*}
    \begin{multicols}{2}
	\includegraphics[width=0.43\textwidth]{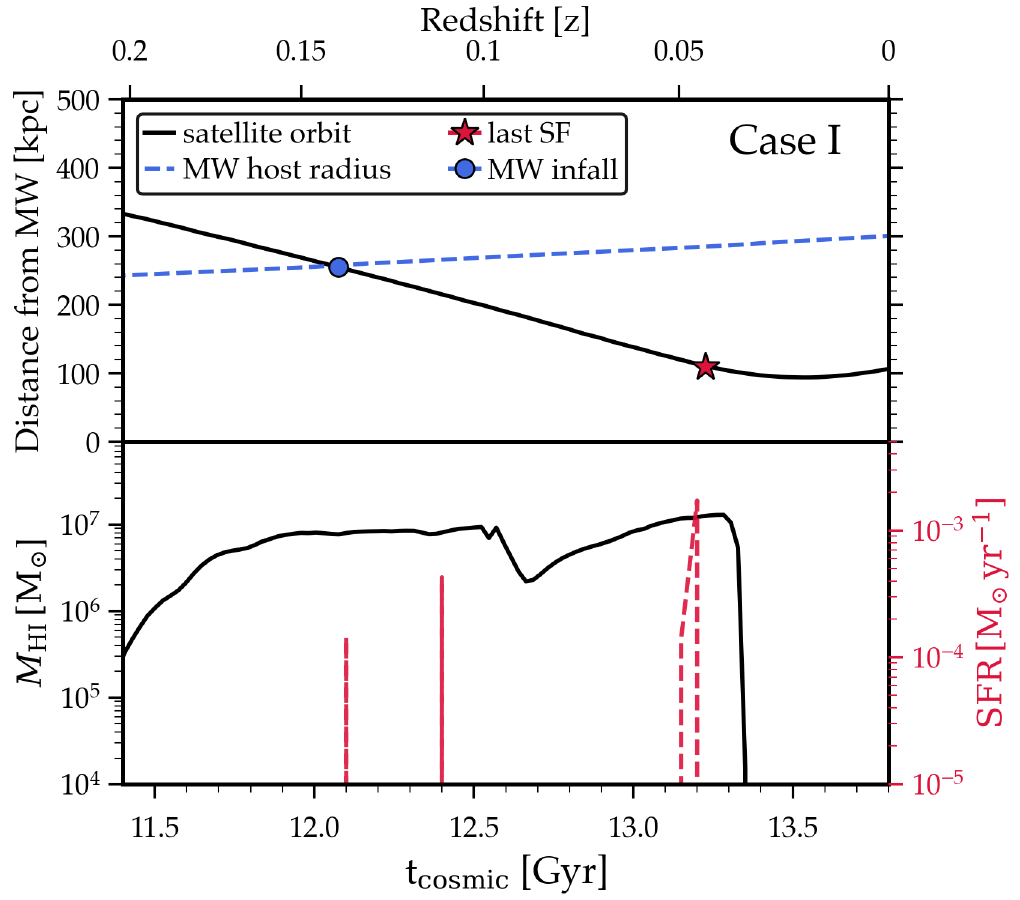}\par
	\includegraphics[width=0.43\textwidth]{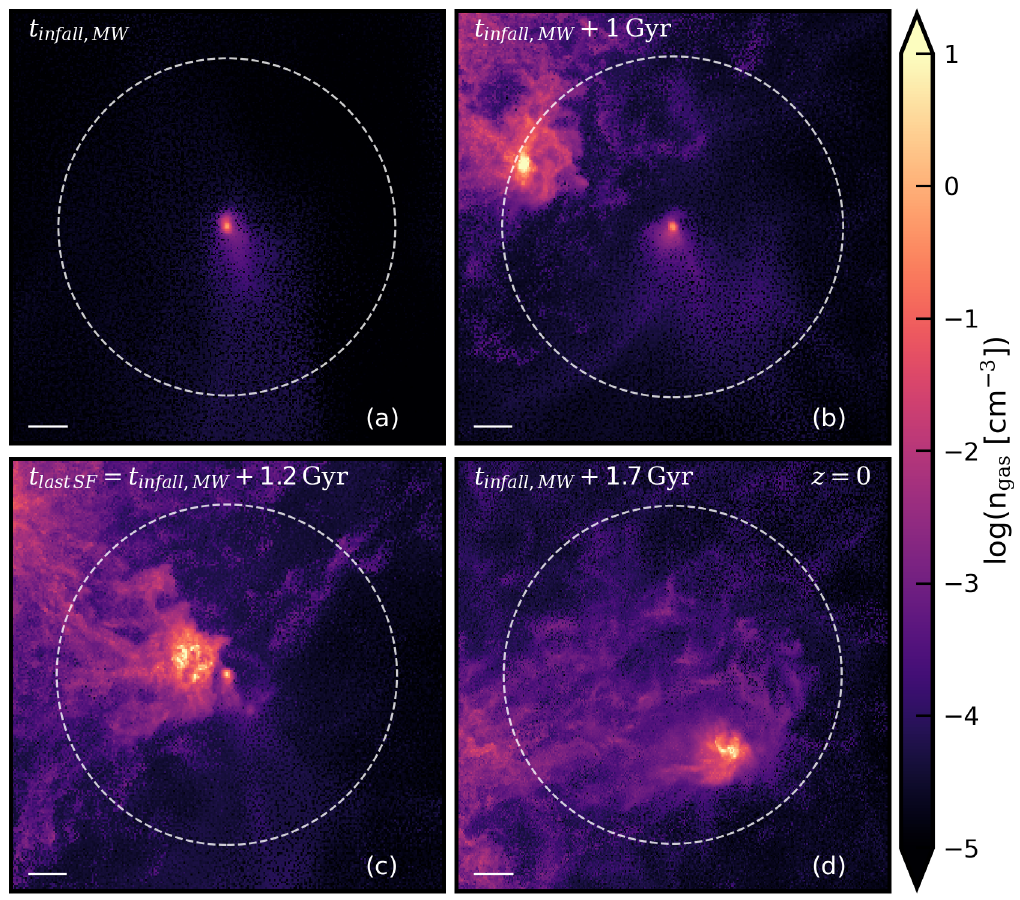}\par
	\end{multicols}
	\vspace{-7 mm}
    \begin{multicols}{2}
	\includegraphics[width=0.43\textwidth]{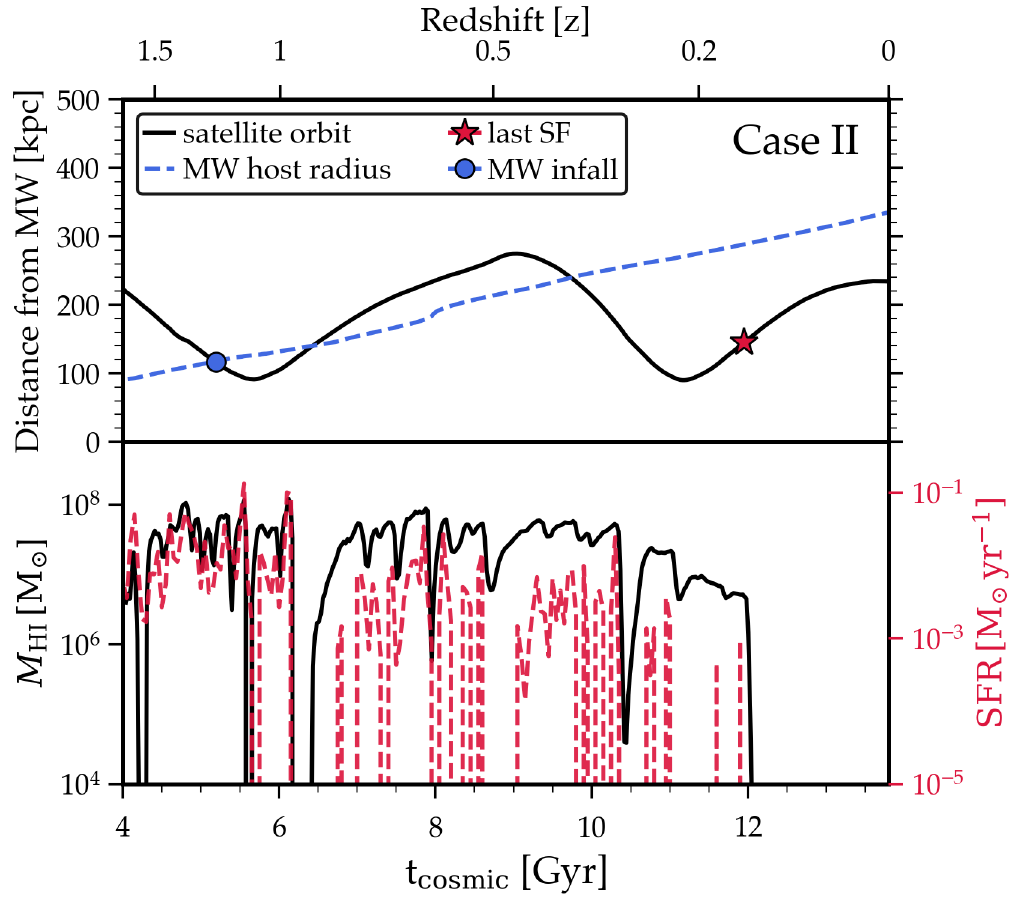}\par
	\includegraphics[width=0.43\textwidth]{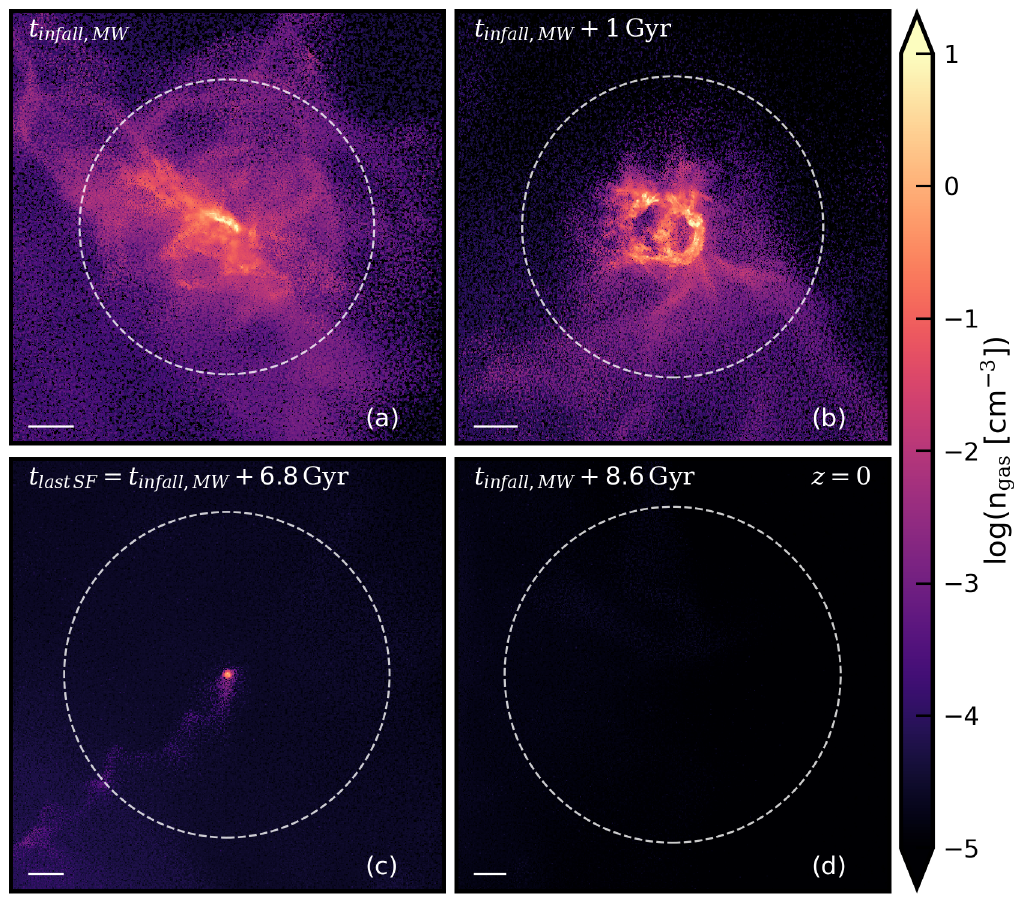}\par
	\end{multicols}
	\vspace{-7 mm}
    \begin{multicols}{2}
	\includegraphics[width=0.43\textwidth]{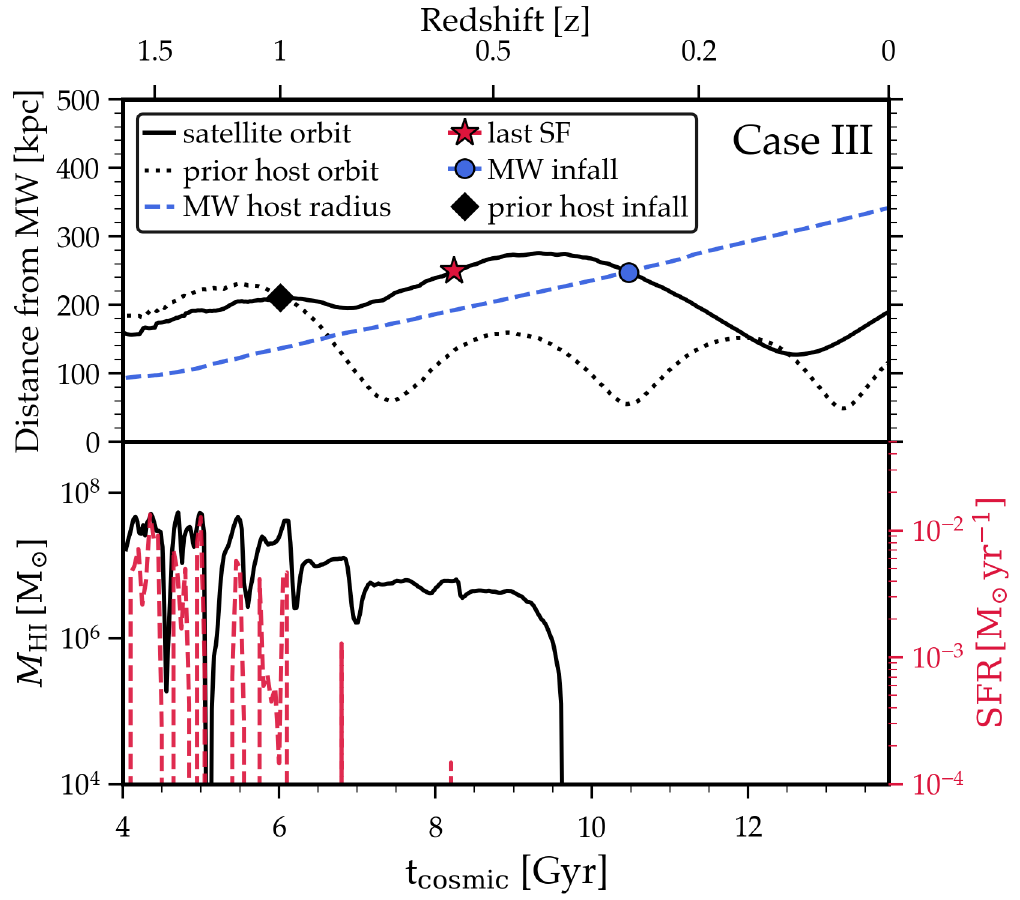}\par
	\includegraphics[width=0.43\textwidth]{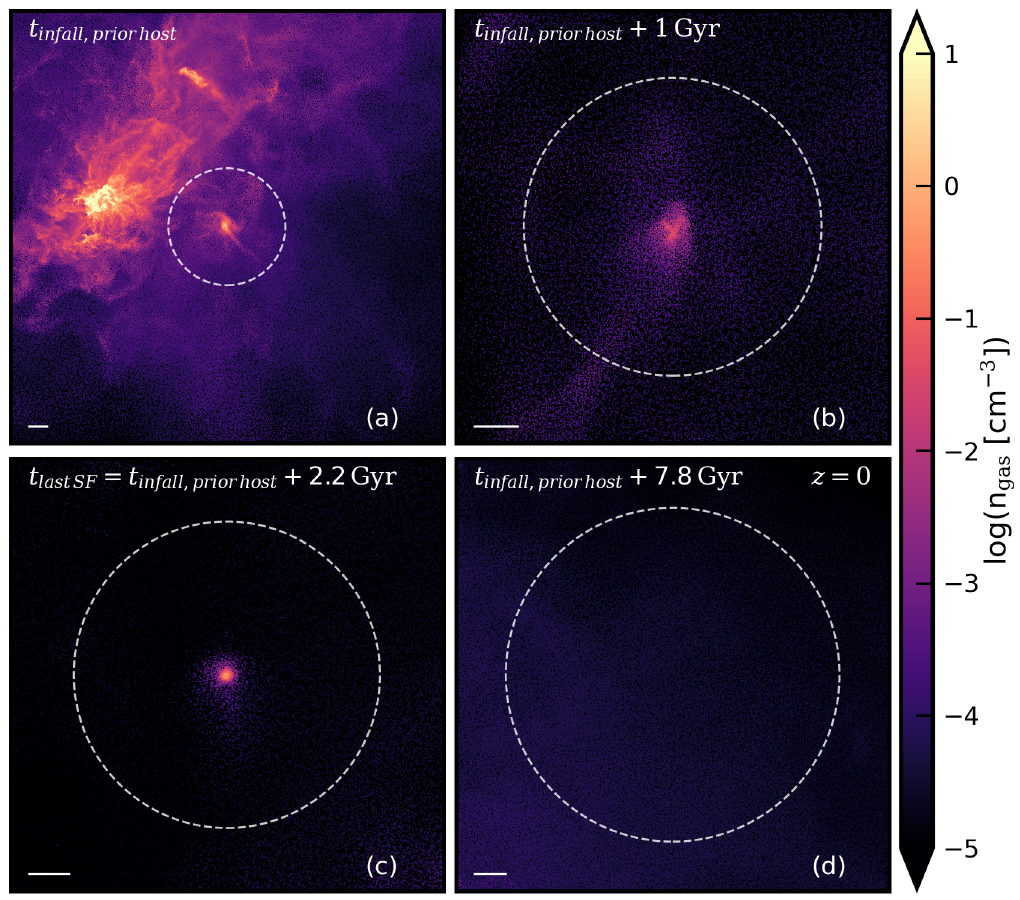}\par
	\end{multicols}
	\vspace{-7 mm}
    \caption{Case studies of quiescent satellite galaxies with representative evolutionary histories.
    From top to bottom: Case I ($\Mstar=6\times10^5\,\Msun$) quenches rapidly after a close fly-by with another satellite and just before first pericentre passage, Case II ($\Mstar=6\times10^7\,\Msun$) quenches after two pericentre passages and has visible gas disturbance following a burst of SF, and Case III ($\Mstar=2\times10^7\,\Msun$) quenches rapidly after momentarily falling into a low-mass group before entering the MW-mass halo.
    \textit{Left:} The orbit, HI mass, and star formation rate (SFR) of each galaxy.
    Each galaxy's SF extends over different amounts of time, and SF typically occurs when $\MHI\gtrsim10^6\,\Msun$. 
    \textit{Right:} Images of gas density across time in each galaxy: (a) at first infall into the MW halo or a low-mass group, (b) 1 Gyr after infall, (c) at the time of the galaxy's last star formation, and (d) at $z=0$.
    Panel (a) is the satellite at the moment marked by a blue dot or black diamond in the left panels, and panel (b) is the satellite at the moment marked by the red star.
    In each panel, the galaxy's velocity towards the center of the MW-mass host halo is in the upward (+y) direction, the bar at lower left is 10 kpc wide, and the dashed circle is the subhalo radius ($\rtwohm$).
    Each galaxy evolves differently, but they all appear to have gas removed from their outskirts first and retain a small dense parcel of gas in their inner regions until quenching.
    }
    \label{fig:case_studies}
\end{figure*}

We discuss three examples of the evolution of quiescent satellite galaxies to help illustrate the general trends we find in the statistical analyses elsewhere in this work.
We highlighted these galaxies with cyan squares in Figures~\ref{fig:QDT} and~\ref{fig:QDT_min}.
Examining the points corresponding to these galaxies on the right panel of Figure~\ref{fig:QDT_min}, we note that each galaxy quenches closest to a different environmental event, which would be easily missed if we only considered quenching delay times with respect to a single event.
Similar to many other satellites, Case I quenches closest to first pericentre passage, but we show here that its evolution is more complicated when we visualize its gas distribution.
Case II quenches closest to second pericentre passage, and represents a relatively common quenching scenario.
Case III is the highest-mass satellite that quenches closest to infall into a low-mass group.
We summarize the properties of these three galaxies in Table~\ref{table:case_studies} and show their evolution in Figure~\ref{fig:case_studies}.

We illustrate the orbit, $\MHI$, and SFR of each case study on the left side of Figure~\ref{fig:case_studies}, and on the right side we visualize the gas density at different times.
We construct projections of gas density by selecting all gas cells within $2-5\times\rtwoh$ of the galaxies. 
We color pixels based on the maximum gas density in each pixel and along the line of sight.
If a pixel is empty, we color it according the the average of the nearest 8 non-empty pixels, to smooth the image slightly.
In each panel on the right, we show the satellite's gas density at four different times: (a) at first infall into the host environment (first crossing of the MW-mass or low-mass host's $\rtwoh$), (b) 1 Gyr after infall, (c) at the time of the satellite's last star formation, and (d) at $z=0$.

First, we consider two satellites (Cases I and II) that have quenched within the MW-mass host environment, without having been in a low-mass group.
These satellites correspond to the blue circles highlighted in Figure~\ref{fig:QDT}.

In the top row of Figure~\ref{fig:case_studies}, we show Case I, an example of gas removal and quenching from ram pressure as a satellite moves through and interacts with the host environment.
Case I falls into the host halo 1.7 Gyr prior to $z=0$ and visually appears to experience gas removal from ram pressure as it first enters the host halo.
The satellite is still able to form stars while experiencing this ram pressure, possibly because while ram pressure removes gas in the outer regions of the galaxy, it can also compress gas in the inner regions to induce star formation \citep{Wright2019,DiCintio2021}.
The last star formation in this satellite occurs at $z=0.04$, $\approx300\, \myr$ prior to its first and only pericentre passage, which brings it within 94 kpc of the host.
It has a quenching delay time of 1.1 Gyr relative to MW infall, consistent with estimates of rapid quenching through ram pressure in a clumpy medium and similar to expectations for LG satellites \citep[e.g.,][]{Fillingham2016}.
Interestingly, based on visual inspection, Case I experiences a close fly-by (within $\approx2\rtwoh$ of Case I) with another satellite galaxy (visible in panels b-d) just as it forms its last star particle.
This signals that satellite-satellite interactions within the MW halo potentially are a significant quenching mechanism, though we leave further exploration of this aspect for future work.

In the middle row of Figure~\ref{fig:case_studies}, we show Case II, a more massive satellite that quenches on a longer timescale relative to MW infall.
The visualization of Case II illustrates how internal stellar feedback can rarefy gas within a galaxy, thereby making the gas more susceptible to ram pressure effects in subsequent orbits within the MW-mass halo.
Such complete gas removal in a satellite with a relatively large halo mass is likely possible because it has been inside the MW-mass host halo experiencing environmental effects for a long time.
The last star formation in Case II happens at $z=0.15$, $\approx750\,\myr$ after the satellite's second pericentre passage, which brings it within 90 kpc of the host.

Case II's long quenching delay time of 6.8 Gyr is consistent with estimates of quenching through gas consumption over an extended time period \citep[e.g.,][]{Fillingham2015,Wetzel2015b}.
However, this satellite shows evidence of both stellar feedback and ram pressure that may contribute to its quenching.
In panel (b) of Figure~\ref{fig:case_studies} (middle row), the gas is first pushed out from the galaxy by supernova feedback following a burst of star formation, which rarefies the gas and spreads it over a larger surface area.
Then at later times in panel (c), only a small dense nugget of gas is left at the center of the satellite with a ram pressure tail of gas extending to the lower left, coinciding with the time of last star formation in this satellite.
Stellar feedback followed by ram pressure likely more efficiently removes gas from and quenches intermediate mass satellites ($\Mstar\sim 10^{6-8}\,\Msun$) than ram pressure alone \citep{Yannick2015,Emerick2016}.
Based on visual inspection, this combination of stellar feedback and ram pressure effects seems to be a common scenario leading up to quenching among satellites with $\Mstar=10^{7-8}\,\Msun$ in our simulations.

We showed in previous sections that some satellites experience significant preprocessing in low-mass groups before entering the MW host halo.
To further explore the effects of group preprocessing, we consider a satellite (Case III) that quenched closest to infall into a prior host, which corresponds to the yellow diamond that we highlighted in Figures~\ref{fig:QDT} and~\ref{fig:QDT_min}.

In the bottom row of Figure~\ref{fig:case_studies}, we show the evolution of Case III, a $\Mstar\approx2\times10^7\,\Msun$ satellite that quenches as a central \textit{before} infall into the MW-mass halo after interacting with a low-mass group.
Its quenching delay time relative to MW infall is -2.2 Gyr, whereas it is +2.2 Gyr relative to infall into a low-mass group.
It is efficiently quenched after a short ($\approx270\,\myr$) encounter with a low-mass group at $z\approx1$ ($\approx4.5\,\gyr$ before MW infall).
During this time, gas is being visibly removed from the galaxy in panel (a), likely because of ram pressure, but also possibly from tidal interactions such as shocking \citep{Marasco2016}.
Case III does not stay bound to its prior host, so they are not near each other at $z=0$, but it continues to lose gas even after this encounter in panel (b), is able to continue forming stars until $z=0.57$ in panel (c), and is completely devoid of gas by $z=0$ in panel (d).

The low-mass host of Case III plunges into the MW-mass host halo immediately after this interaction at $z=0.85$.
Case III's low-mass host quenches 5.5 Gyr after infall into the MW-mass halo after experiencing three pericentre passages near $\approx50\,\kpc$ (black diamond in right panel of Figure~\ref{fig:QDT}).
The dark matter halo of the low-mass host undergoes significant mass-loss during this time, bringing it from a peak subhalo mass of $\Mpeak\approx2.5\times10^{10}\,\Msun$ to $\Mtwohm(z=0)\approx7.5\times10^8\,\Msun$, which is comparable to its SMC-like stellar mass of $\Mstar(z=0)\approx1.3\times10^8\,\Msun$.
The low-mass host galaxy contains no gas at $z=0$, and serves as an additional example of how satellites with $\Mstar>10^8\,\Msun$ can also be quenched by the host environment, given multiple close pericentre passages, but still survive as an intact satellite galaxy to $z=0$.

Though the evolutionary histories of our three case studies differ significantly, they all visually appear to lose gas in their outer regions while retaining a small dense inner region of gas that is able to continue forming stars for up to $\approx7\,\gyr$ after infall into the MW halo.
This is consistent with the results of \citet{Hausammann2019}, who analyzed wind-tunnel simulations of ram pressure effects on galaxy formation, and found that the extended hot gas in low-mass galaxies ($\Mstar\approx10^{6-7}$) is removed on short timescales while the central cold gas can persist and continue to form stars.
Furthermore, \citet{DiCintio2021} also found that if simulated LG satellites contain enough cold gas around the time of infall, then pericentre passages can actually promote star formation by compressing gas in the inner regions of the satellite.
Taken together with our results, this could mean that the rapid quenching of satellites at $\Mstar\approx10^{6-7}\,\Msun$ in our simulations upon first infall may be because of increased effectiveness of ram pressure at low galaxy masses.

\subsection{Comparison to other simulations}\label{sec:qf_compare}

\begin{figure}
	\includegraphics[width=\columnwidth]{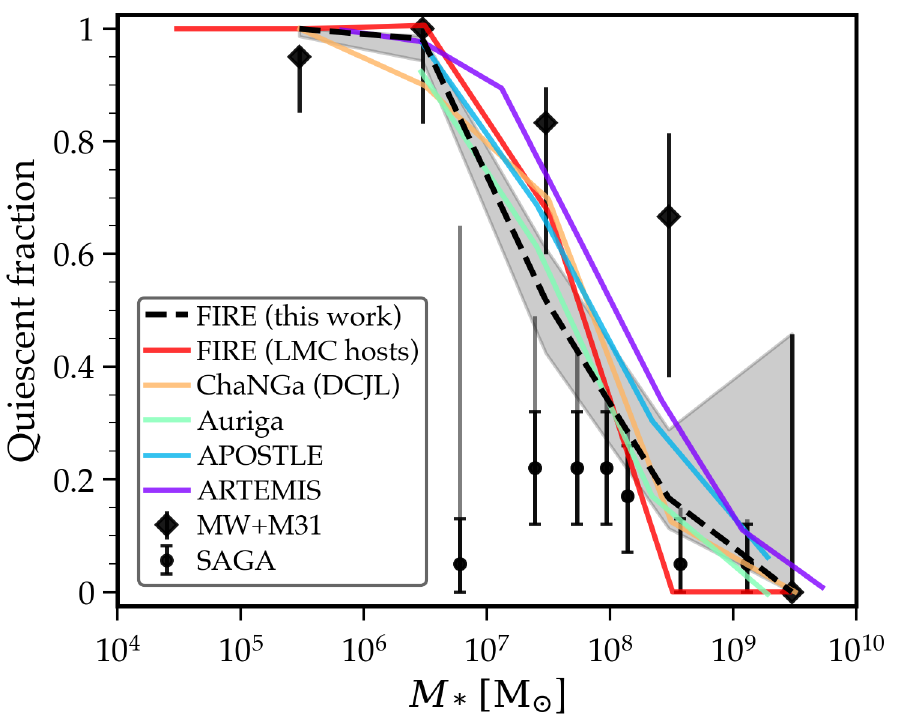}\par
    \vspace{-2 mm}
    \caption{A comparison of satellite quiescent fractions from different cosmological zoom-in simulations and observations.
    All simulations are of MW-mass hosts, except the red line, which shows results for LMC-mass hosts from the FIRE-2 simulations.
    We use our fiducial definition of quiescent for the dashed black line and grey shaded region: no star formation in a galaxy within the last 200 Myr and $\MHI<10^6\,\Msun$ at $z=0$.
    The general agreement both amongst simulations and between simulations and the LG is noteworthy, with the scatter amongst the different simulation means being much smaller than the host-to-host scatter of our simulations (see Figure~\ref{fig:qf_obs}).
    In comparison, the SAGA survey's turnover below $\Mstar\approx10^{7.5}\,\Msun$ is conspicuous and could indicate physically different host environments and/or observational incompleteness.
    Note that although we omit uncertainties for the other simulations for clarity, there is significant overlap between the simulations' uncertainties and that of the observations.
    }
    \label{fig:QF_compare}
\end{figure}

In Figure~\ref{fig:QF_compare} we show the quiescent fraction of satellites across all of our simulations and 1-sigma statistical scatter (rather than the much larger host-to-host scatter we show in Figure~\ref{fig:qf_obs}) compared to observations and results from five other suites of cosmological zoom-in simulations.
We note that the black dashed line and shaded region are the same as the solid black line and shaded region from Figure~\ref{fig:qf_isolated}.
Our simulation quiescent fraction is similar to the LG across all satellite stellar masses, though they are not necessarily consistent at the 1-sigma level.
Interestingly, the simulations are consistent with SAGA at the 1-sigma level at $\Mstar\gtrsim10^8\,\Msun$, where the LG quiescent fraction is actually a bit higher than the simulations, though the LG suffers from small sample size here.
In contrast to the simulations, the SAGA survey has a noticeable turnover around $\Mstar\lesssim10^{7.5}\,\Msun$.
The results from the other simulation groups show similar separations from the SAGA survey at low satellite masses, which have often been used to claim a tension between simulations and SAGA.
Below, we review our results as compared to other simulations and further discuss how they compare to observations of satellites in the LG and the SAGA survey.

\citet{Jahn2021} examined the quiescent fraction of satellite galaxies in FIRE-2 simulations using satellites around six isolated LMC-mass galaxies ($\Mtwohm\approx10^{11}\,\Msun$, $\rtwoh\approx170-210\,\kpc$).
These simulations have resolutions ($\rm{m}_{\rm baryon,ini}=880-7070\,\Msun$) comparable to or better than the simulations we use ($\rm{m}_{\rm baryon,ini}=3500-7070\,\Msun$), so they were able to analyze satellites down to $\Mstar=10^4\,\Msun$.
\citet{Jahn2021} used a time-based criterion to define quiescent satellites in their sample, choosing a conservative lookback time to last star formation of 500 Myr.
They compared quiescent fractions around their simulated LMC-mass galaxies to LG observations and to results from the simulations used in this work in the middle panel of their Figure 3.
Note that in their comparison, \citet{Jahn2021} uniformly applied the lookback time of 500 Myr as the criterion for quiescence to all FIRE-2 simulations for consistency, which is slightly different from the results we present using our fiducial definition of quiescence (Section~\ref{sec:q_def}).

Comparing FIRE-2 satellites in Figure~\ref{fig:QF_compare}, we find that the results for satellites of LMC-mass galaxies (red line) are remarkably similar to satellites of MW-mass galaxies (black dashed line).
The quiescent fraction of satellites for LMC-mass and MW-mass hosts only differ by about $\pm0.2$ at $\Mstar=10^{7-9}\,\Msun$.
Furthermore, \citet{Jahn2021} find evidence for environmental quenching based on star formation in satellites shutting off within $\lesssim2-4\,\gyr$ of infall into and within distances of $2R_{\rm 200m}$ of their LMC-mass hosts.
This is similar to results from \citet{Fillingham2015} and \citet{Wetzel2015a}, who used the dark matter-only ELVIS suite to determine quenching timescales for satellite galaxies in the LG.
These authors predicted that satellites with $\Mstar\approx10^{6-8}\,\Msun$ are environmentally quenched on relatively short timescales of $\approx2\,\gyr$ from infall, whereas satellites with $\Mstar\approx10^{8-11}\,\Msun$, experience a prolonged gas consumption mode of quenching on longer timescales of $\approx8\,\gyr$.

The LMC hosts are, on average, five times less massive than the low-mass MW hosts, which could mean that there is little dependence of environmental quenching of satellites on host halo masses between $\Mtwohm\approx10^{11-12}\,\Msun$.
Similar to MW-mass galaxies in FIRE-2, the LMC-mass galaxies also tend to have hot virialized gas in their outer haloes \citep{Stern2021}, which may be responsible for environmentally quenching satellites.
Many of the surviving quenched LMC satellites have orbited within small distances ($\approx10-20\,\kpc$) of the host \citep{Jahn2021}, where the volume tends to be filled with cold, dense gas \citep{Stern2021} that can efficiently remove gas from and quench satellites.
Whereas, surviving satellites in the MW simulations tend to have larger pericentre distances of $\approx50-100\,\kpc$ (Santistevan et al. in prep), and most of the cold dense gas in the inner regions of these simulations is confined to the small volume of the MW's disk.
However, a direct comparison of the ram pressure experienced by quiescent satellites around MW and LMC hosts is needed to fully explain the similarity in their quiescent fractions, which we defer to future work.

\citet{Akins2021} investigated quenching of satellite galaxies ($\mstarlowlim$) in four simulations of MW-like galaxies from the ChaNGa DC Justice League suite (DCJL).
They used a specific star formation rate (sSFR) of $<10^{-11}\,\rm{yr}^{-1}$ averaged over the last 100 Myr to define quiescent galaxies.
\citet{Akins2021} found broad agreement with the quiescent fraction in the LG and some tension with the quiescent fraction from SAGA.
They also found a positive trend in quenching delay time (relative to MW infall) versus stellar mass, similar to our Figure~\ref{fig:QDT} (right), that indicates strong environmental effects for intermediate-mass satellites ($\Mstar=10^{6-8}\,\Msun$).
\citet{Applebaum2021} introduced higher resolution versions of the DCJL simulations, and focused on the quenching of satellite UFDs. 
UFDs are almost uniformly quenched early by reionization in these simulations, but there are a few cases of gas-rich UFDs at $z=0$.

\citet{Karunakaran2021} re-calculated the quiescent fraction of satellites in the SAGA survey using archival GALEX UV imaging, and compared their results to satellites from the APOSTLE and Auriga simulations (originally presented in \citealt{Simpson2018}).
Their definition of a star-forming simulated satellite galaxy ($\rm{SFR}_{\rm sim}>0\, \Msun\rm{yr}^{-1}$) is based on scaling the HI content of satellites to star-formation relations for their simulations found in prior work.
Results from the APOSTLE and Auriga simulations broadly agree with the LG, the results of this work, and the simulations discussed above.
In their observational re-analysis, the GALEX-derived quiescent fraction for SAGA is consistent with the original quiescent fraction reported by the SAGA team.
This led \citet{Karunakaran2021} to claim a tension between SAGA and the APOSTLE and Auriga simulations, and they posited that an explanation for this tension could lie in physical differences in satellite evolution around SAGA hosts as compared to the LG, which we also find some evidence for in Section~\ref{sec:qf_host_env}.

Using the ARTEMIS simulations, \citet{Font2022} found a quiescent fraction similar to the LG and other simulations.
They defined a galaxy as quiescent if its instantaneous specific star formation rate is zero at $z=0$, and they verified that this agrees with lookback time and time-averaged sSFR criteria.
However, \citet{Font2022} also showed that by including a surface brightness limit ($\mu_{e,r}\approx25\,\rm{mag}\,\rm{arcsec}^{-2}$) in their comparisons to the SAGA survey, the ARTEMIS simulations can reproduce the quiescent fraction of satellites observed in SAGA.
Therefore, they suggest that the reported tensions between simulations and SAGA exist because the simulations do not account for the potential implicit observational selection of higher surface brightness satellites in the SAGA survey.
Though we do not perform the same correction for observational completeness, we do find some evidence for better agreement between our simulations and the SAGA survey when we measure the quiescent fraction in projection (Section~\ref{sec:qf_obs}).

It is clear from Figure~\ref{fig:QF_compare} that statistical results from different zoom-in simulation groups agree to better than our host-to-host scatter, and comparable to (or better than) observational uncertainties in satellite quiescent fraction, despite differences in subgrid physics, numerical implementation, resolution, and definition of quiescence.
This suggests that the details of satellite quenching may in fact not be that sensitive to subtleties in hydrodynamic method and modeling of the ISM and feedback.
However, all of the simulations discussed above used broadly similar mass resolutions, so more work is needed to test the robustness of these results at much higher resolution.


\section{Summary and Discussion}\label{sec:discussion}

We have presented an overview of the environmental quenching of $z=0$ galaxies ($\Mstar=10^{5-10}\,\Msun$) around 14 MW/M31-mass host galaxies from the FIRE-2 simulations.
We found that most of the simulated satellite galaxies ($\dlim$) are gas-poor and quiescent, similar to Local Group (LG) satellites.
We have analyzed broad characteristics of the host environment that may be responsible for environmental quenching, and how events like infall, pericentre passage, and group preprocessing coincide with quenching.
We found evidence for mass-dependence in environmental quenching of satellite galaxies with respect to both satellite and host mass.
We list and discuss our main conclusions in the following subsections.

\subsection{Host environment trends}

\begin{enumerate}
    \item The quiescent fraction of galaxies increases as distance from a MW-mass host decreases (Figures~\ref{fig:HI_content} and~\ref{fig:qf_vs_dist}).
    \item Generally, lower-mass satellites ($\Mstar\lesssim10^7\,\Msun$) are quiescent, higher-mass satellites ($\Mstar\gtrsim10^8\,\Msun$) are star-forming, and about half of intermediate mass satellites ($\Mstar\approx10^{7-8}\,\Msun$) are quiescent (Figures~\ref{fig:qf_isolated} and~\ref{fig:qf_obs}).
    \item Hosts with more massive CGM tend to have a higher quiescent fraction of massive satellites at $\Mstar=10^{8-9}\,\Msun$ (Figure~\ref{fig:QF_hostenv}, left).
    \item We find no obvious significant differences in the quiescent fraction of satellites around isolated and paired (LG-like) hosts, but we discuss some qualifications below. (Figure~\ref{fig:QF_hostenv}, right).
    \item In the FIRE-2 simulations, the quiescent fraction of satellites around MW-mass hosts is similar to the quiescent fraction of satellites around LMC-mass hosts (Figure~\ref{fig:QF_compare}).
\end{enumerate}

The MW-mass host environment effectively environmentally quenches satellite galaxies, but trends in quenching with particular host properties are not overwhelming.
We note that though the trend we find with host CGM mass (and to a lesser extent total mass) can explain some of the large host-to-host scatter in the quiescent fraction of massive satellites, they do not provide an explanation for the large scatter at $\Mstar=10^{7-8}\,\Msun$.
The moderate trends we find in the quiescent fraction with host mass and lack of a strong trend with paired versus isolated host environment may hint at observational differences between the SAGA survey and the LG, which we discuss further in Section~\ref{sec:obs_discussion}.

\subsection{Group preprocessing and quenching timescales}

\begin{enumerate}
    \item Satellites with $\Mstar=10^{7-8}\,\Msun$ that were in a low-mass group before falling into the MW-mass halo are more likely to be quiescent compared to satellites without prior group associations (Figure~\ref{fig:GP_qf}). 
    \item About 10 per cent of satellites at $\Mstar<10^7\,\Msun$ quenched \textit{within} their prior host's halo (Figure~\ref{fig:quench_location}).
    \item Quenching delay time relative to first infall (into a low-mass group or a MW-mass halo) positively correlates with satellite stellar mass, such that more massive galaxies take longer to quench after infall (Figure~\ref{fig:QDT}, left). 
    \item Satellites with a prior group association typically quenched faster within the MW halo than satellites without prior group associations (Figure~\ref{fig:QDT}, right).
    \item Lower-mass satellites tend to quench as centrals or upon first infall into any host. More massive quiescent satellites have typically experienced $\geq1-2$ pericentre passages and quench after $\approx2.5-5\,\gyr$ within the MW halo (Figures~\ref{fig:QDT},~\ref{fig:QF_peris}, and~\ref{fig:QDT_min}).
\end{enumerate}

Group preprocessing, or environmental quenching in low-mass groups before MW infall, is an effective mechanism in the quenching of MW satellites in our simulations.
This agrees with previous studies that have shown that interactions between low-mass galaxies are capable of disturbing their gas, such that it may be more easily removed by ram pressure within the halo of a massive host \citep{Marasco2016,Pearson2018}.
We find that many, especially lower-mass, satellites quench closest to infall into a low-mass group, before entering the MW halo.
This is consistent with results from \citet{Wetzel2015b} who found that lower-mass subhaloes were more likely to experience group preprocessing.
These effects could be especially important for satellite quenching in the LG, where the LMC may have brought in as many as three of its own satellites ($\mstarlowlim$) and even more UFDs \citep{Jahn2019}.
Similarly, satellite-satellite interactions within the MW halo may also contribute to quenching.

\subsection{Comparisons to observations and other simulations}\label{sec:obs_discussion}

\begin{enumerate}
    \item The simulations reproduce the average trend observed in the Local Group wherein the quiescent fraction of satellites increases as satellite stellar mass decreases over $M_{*}=10^{5-10}\,\rm{M}_{\odot}$ (Figure~\ref{fig:qf_obs}). Though the simulations are consistent with the SAGA survey’s quiescent fraction at $M_{*}\gtrsim10^8\,\rm{M}_{\odot}$, the simulations do not generally reproduce SAGA’s turnover to low quiescent fractions at $M_{*}\lesssim10^{7.5}\,\rm{M}_{\odot}$ (Figure~\ref{fig:QF_compare}).
    \item The quiescent fraction of satellites in FIRE-2 and results from several other zoom-in simulation groups agree remarkably well with each other, despite underlying differences in simulations and analyses (Figure~\ref{fig:QF_compare}).
\end{enumerate}

It is difficult to identify a single reason why the LG and SAGA survey quiescent fractions lie at opposite ends of our simulation host-to-host scatter.
Based on the similarities between quiescent fractions around the isolated and paired hosts in our simulations, SAGA's selection of isolated MW-mass hosts cannot explain the large difference between the MW-M31 pair in the LG and SAGA hosts.
We find that simulated MW-mass hosts with lower CGM mass have lower quiescent fractions of massive satellites, similar to SAGA, but this cannot account for the difference between the LG and SAGA at lower satellite masses.
Furthermore, when we allow foreground and background non-satellites to scatter into a mock-SAGA survey footprint, the distribution of quiescent fractions for lower-mass satellites extends to lower values, indicating that non-satellite interlopers in SAGA may also bias the quiescent fraction to lower values.

Moreover, infall into any host halo and pericentre passage(s) within a MW-mass halo seem to be the dominant events in the environmental quenching of satellites in the simulations.
If observed LG satellites fell in earlier than SAGA satellites, similar to the satellites of paired versus isolated hosts in the simulations, such a bias in satellite accretion history could be a reason for their disparate quiescent fractions, as LG satellites would have experienced environmental effects for longer times.
Overall, the tension between the quiescent fraction of satellites in the LG and the SAGA survey could be due to a combination of differences in host environments (host CGM/total mass), satellite accretion times, and observational selection, but none of these is able to fully explain the tension on its own to the extent that we have explored in our analysis.

\subsection{Caveats}

Resolution plays an important role in how well we are able to model the processes of environmental quenching.
Though the internal dynamics of galaxies at $\Mstar\gtrsim10^6\,\Msun$ are likely sufficiently resolved for our analysis, the less massive galaxies likely suffer from over-quenching because of bursty star formation and stellar feedback that too rapidly remove their gas reservoirs.
\citet{Hopkins2018} found that resolution can significantly alter the SFR and stellar mass of isolated low-mass galaxies in FIRE.
In particular, higher-resolution ($m_{\rm baryon}=500-880\,\Msun$) isolated galaxies ($\Mstar=10^{5-6}\,\Msun$) in the FIRE-2 simulations generally remain star-forming to $z=0$ \citep[e.g.,][]{Fitts2017}.
This is partly because of their isolated environment, but in comparison there are probably resolution issues at this mass scale in our galaxy sample that are responsible for their quenching and which we will explore further in future work.

A more nuanced analysis of the gaseous haloes (CGM) around host galaxies is also needed to determine exact environmental trends and whether ram pressure alone is responsible for quenching satellites, or if there are more processes involved such as tidal interactions.
While we have seen visual evidence of gas removed by ram pressure and internal stellar feedback rarefying satellite gas, we have not yet accounted for these effects quantitatively.
Resolution in the CGM can be important for fully understanding quenching in the context of ram pressure too \citep{Simons2020}, though we have not evaluated this effect here.

Certain choices in our analysis methodology and in the physics of the simulations we used may somewhat narrow the scope of our results.
We have examined only the surviving population of satellites at $z=0$, but the explicit inclusion of recently disrupted satellite galaxies and separation of splashback galaxies in our analysis would provide important context for galaxy evolution at higher redshift, especially with respect to the effects of reionization.
We find that the quiescent fraction of low-mass galaxies is elevated out to $\approx1\,\mpc$ or a few times the halo radius of our MW-mass hosts. 
Such possible signals of environmental quenching outside of the immediate host environment have been noted by other authors, but we do not explore them further here \citep{Bahe2013,Applebaum2021,Pasha2022,Yang2022}.
The MW simulations we use from the FIRE-2 simulations do not include any form of AGN feedback, but we note that this feedback channel does not seem to be necessary to reproduce the quiescent fraction of observed LG and SAGA survey satellite galaxies.
In future work, we will also consider the effects of different physics, namely the inclusion of cosmic rays and magnetohydrodynamics, on host CGM and satellite quenching.

\section*{Acknowledgements}

We thank the anonymous reviewer for their careful reading of our manuscript and their many insightful comments that improved the quality of this paper.

JS was supported by an NSF Astronomy and Astrophysics Postdoctoral Fellowship under award AST-2102729.
JS and AW received support from: the NSF via CAREER award AST-2045928 and grant AST-2107772; NASA ATP grants 80NSSC18K1097 and 80NSSC20K0513; HST grants AR-15057, AR-15809, GO-15902, GO-16273 from the Space Telescope Science Institute (STScI), which is operated by the Association of Universities for Research in Astronomy, Inc., for NASA, under contract NAS5-26555; a Scialog Award from the Heising-Simons Foundation; and a Hellman Fellowship.
IBS received support from NASA, through FINESST grant 80NSSC21K1845.
Support for JM is provided by the NSF (AST Award Number 1516374), the Grace Steele Foundation, and Downing College.
MBK acknowledges support from NSF CAREER award AST-1752913, NSF grants AST-1910346 and AST-2108962, NASA grant NNX17AG29G, and HST-AR-15006, HST-AR-15809, HST-GO-15658, HST-GO-15901, HST-GO-15902, HST-AR-16159, and HST-GO-16226 from STScI.
BP received support from the REU program at UC Davis through NSF grant PHY-1852581.
We performed this work in part at the Aspen Center for Physics, supported by NSF grant PHY-1607611, and at the KITP, supported NSF grant PHY-1748958.

We ran simulations using the Extreme Science and Engineering Discovery Environment (XSEDE), supported by NSF grant ACI-1548562; Frontera allocations AST21010 and AST20016, supported by the NSF and TACC; Blue Waters via allocation PRAC NSF.1713353 supported by the NSF; the NASA HEC Program through the NAS Division at Ames Research Center.
Some computations in this paper were run on the Odyssey cluster supported by the FAS Division of Science, Research Computing Group at Harvard University.
We gratefully acknowledge use of the IPython package \citep{ipython}, NumPy \citep{numpy}, SciPy \citep{scipy}, Numba \citep{numba}, matplotlib \citep{matplotlib}, and the WebPlotDigitizer tool \citep{webplotdigitizer}.

This research made use of data from the SAGA Survey \url{sagasurvey.org}. 
The SAGA Survey was supported by NSF collaborative grants AST-1517148 and AST-1517422 and by Heising–Simons Foundation grant 2019-1402.

This work was conducted on Tonkawa Indigenous land and Patwin Indigenous land.

\section*{Data Availability}

The FIRE-2 simulations are publicly available \citep{Wetzel2022} at \url{http://flathub.flatironinstitute.org/fire}.
Additional FIRE simulation data is available at \url{https://fire.northwestern.edu/data}.
A public version of the \textsc{Gizmo} code is available at \url{http://www.tapir.caltech.edu/~phopkins/Site/GIZMO.html}.
The publicly available software packages used to analyze these data are available at: \url{https://bitbucket.org/awetzel/gizmo\_analysis}, \url{https://bitbucket.org/awetzel/halo\_analysis}, and \url{https://bitbucket.org/awetzel/utilities} \citep{WetzelGizmoAnalysis2020,WetzelHaloAnalysis2020}.



\bibliographystyle{mnras}
\bibliography{ref}



\appendix

\section{Definition of quiescence}
\label{sec:appendix_q_def}

\begin{figure}
    \centering
    \includegraphics[width=\columnwidth]{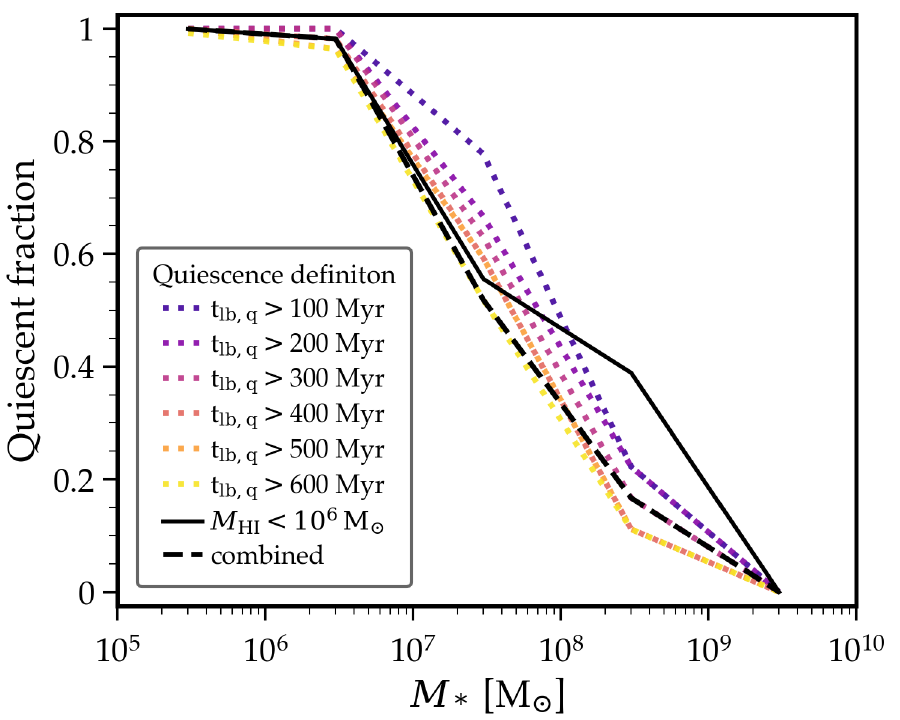}
    \vspace{-6 mm}
    \caption{The effects of varying criteria in our definition of quiescence, compared to our fiducial `combined' definition (no star formation in a galaxy within the last 200 Myr and $M_{\rm HI}<10^6\,\Msun$ at $z=0$).
    We stack all satellites together instead of showing a host-to-host mean.
    There are some small differences in the quiescent fraction when we use lookback times of $100-600\,\myr$ to last star formation.
    An alternative definition of quiescence based on the mass of HI in a galaxy yields identical results for $M_{\rm HI}<10^{4-7}\,\Msun$, though this definition has a higher quiescent fraction at $\Mstar=10^{8-9}\,\Msun$ than the time-based definitions.
    Our combined definition takes into account both recent star formation and the potential for further star formation with HI, which gives a more conservative definition of quiescence that is less sensitive to time.
    }
    \label{fig:QF_vary_time}
\end{figure}

Here we test the effects of varying the choice of lookback time to last star formation and/or $M_{\rm HI}$ in our definition of quiescent.
We use six different times spaced by 100 Myr, and we show the quiescent fraction using these times for all satellites stacked together in Figure~\ref{fig:QF_vary_time}, where we compute the fraction of all satellites in each bin rather than a mean over the values for each host.
All times indicate that the satellite population at $z=0$ is almost completely quiescent at $\Mstar=10^{5-7}\,\Msun$ and mostly star-forming at $\Mstar=10^{8-10}\,\Msun$.
However, there are variations in quiescent fraction amongst the different times, particularly at $\Mstar=10^{7-8}\,\Msun$ where the scatter is about $0.1-0.2$.
The longer 600 Myr time could be missing some satellites that have only recently been accreted and quenched by the host environment, and the shorter 100 Myr time is so short that it unfairly labels satellites as quiescent that have not recently formed stars but likely will in the near future.

We also consider an alternative definition of quiescence based on the absence of significant amounts of HI gas.
We explored different mass thresholds ranging from $M_{\rm HI}=10^{4-9}\,\Msun$, which resulted in identical quiescent fractions for $M_{\rm HI}<10^{4-7}\,\Msun$, and unphysically quiescent populations for higher thresholds.
The stability of the quiescent fraction over a large range of HI mass thresholds is because of a lack of galaxies in our sample with $M_{\rm HI}\lesssim10^6\,\Msun$ (see Figure~\ref{fig:HI_content}).
Within the host halo ($\approx300-400\,\kpc$) this is typically interpreted as highly effective environmental gas removal, but the fact that the trend persists out to around 2 Mpc could indicate resolution or other numerical effects that prohibit galaxies from retaining smaller quantities of HI.

\begin{figure}
    \centering
    \includegraphics[width=\columnwidth]{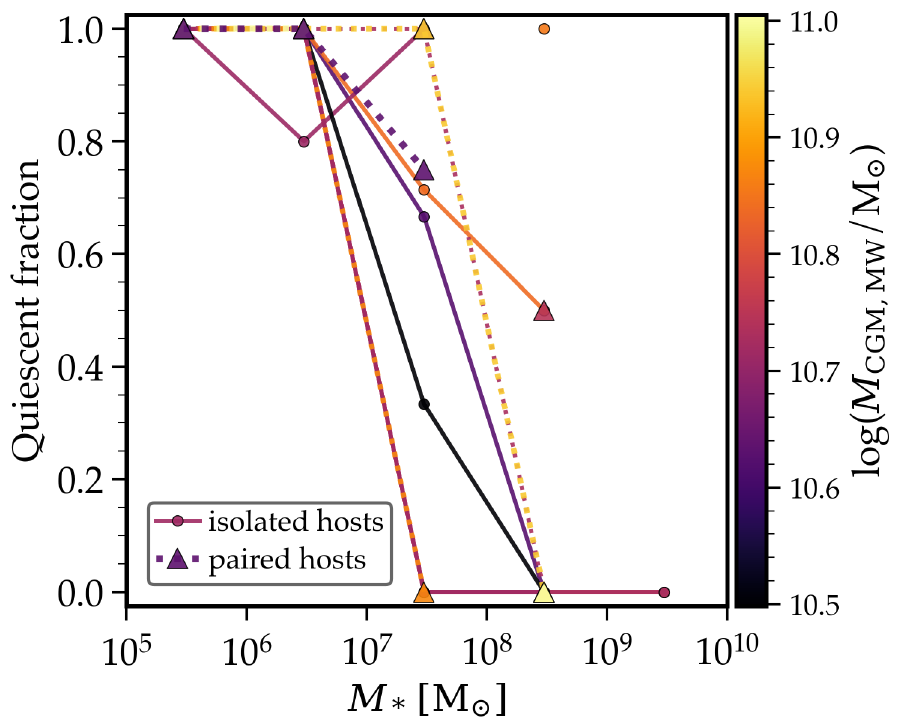}
    \vspace{-6 mm}
    \caption{The quiescent fraction of satellite galaxies around each of the 14 hosts in the simulations using our fiducial `combined' definition of quiescence.
    We color individual lines by the mass of the host CGM.
    Some hosts lack any satellites in one or more mass bins at $\Mstar>10^7\,\Msun$, which causes their points to be disconnected or the line to be truncated at high masses.
    Many hosts lie at the upper or lower extremes of the full host set, and the host-to-host scatter is particularly large at $\Mstar=10^{7-8}\,\Msun$.
    }
    \label{fig:QF_each_host}
\end{figure}

We opt to combine the time and $\MHI$ criteria to create a conservative definition of quiescence that takes into account numerical resolution: no star formation in a galaxy within the last 200 and $M_{\rm HI}<10^6\,\Msun$.
We find that when we change the time in this combined definition, the corresponding variations in the quiescent fraction are small ($\lesssim0.05$) and there is no longer a strong time dependence to our results where we subsample host systems or satellite populations.
We show the quiescent fraction around each host in the simulations using this definition in Figure~\ref{fig:QF_each_host}.

We also briefly explore other definitions of quiescence used in the works that we compare to in Section~\ref{sec:qf_compare}.
For example, \citet{Akins2021} used a specific star formation rate (sSFR) of $<10^{-11}\,\rm{yr}^-1$ averaged over the last 100 Myr.
We note that by this definition the formation of a single star particle in our simulations over the last 100 Myr would yield sSFR$\geq10^{-11}\,\rm{yr}^-1$ (star-forming) in a galaxies of $\Mstar\leq7\times10^6\,\Msun$ and sSFR$<10^{-11}\,\rm{yr}^-1$ (quiescent) in more massive galaxies.
However, the time part of our definition of a quiescent galaxy requires that the sSFR over the last 100 Myr be zero regardless of galaxy mass and could lead us to measure lower quiescent fractions at high masses than \citet{Akins2021}, though this does not appear to lead to much difference in Figure~\ref{fig:QF_compare}.
\citet{Karunakaran2021} defined a star-forming satellite galaxy by $\rm{SFR}_{\rm sim}>0\, \Msun\rm{yr}^{-1}$ based on scaling the HI content of satellites to star-formation relations for their simulations.
This is similar to our time criterion and we incorporate an HI criterion in our definition of quiescence.
\citet{Karunakaran2021} also report qualitative agreement with the definition of \citet{Akins2021}.
Furthermore, \citet{Font2022} also define a galaxy as quiescent if its instantaneous specific star formation rate is zero, and they verify that this agrees with a time-based criterion and a time-averaged sSFR like that used in \citet{Akins2021}.
Thus, any small differences amongst the different definitions are unlikely to yield qualitatively different results for the quiescent fraction of satellites.

\section{Quenching delay time with respect to first pericentre passage}\label{sec:appendix_QDT}

\begin{figure}
	\includegraphics[width=0.47\textwidth]{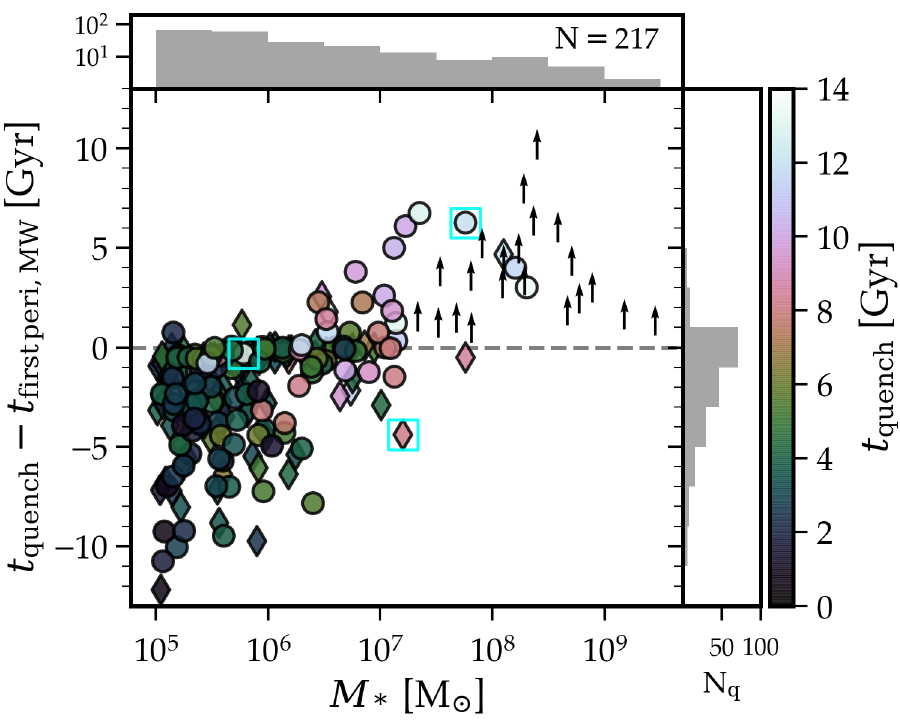}\par
    \vspace{-2 mm}
    \caption{Quenching delay time versus stellar mass for all satellites at $z=0$, relative to first pericentre passage within a MW-mass halo.
    Quiescent satellites are filled symbols, colored according to the cosmic time at quenching, and other conventions are the same as in Figure~\ref{fig:QDT}.
    Histogram at the top shows the stellar mass distribution of all satellites.
    Many quiescent satellites have a quenching delay time near zero, but they do not cluster as strongly near zero compared to first infall.
    }
    \label{fig:QDT_firstperi}
\end{figure}

Figure~\ref{fig:QDT_firstperi} shows quenching delay time with respect to the timing of first pericentre passage around a MW-mass host.
If a satellite is on its first infall and has not yet experienced a pericentre passage around the MW-mass host (23 out of 240 satellites are on such a first infall), then we exclude it from this figure.
Surprisingly, this version of quenching delay time does not yield a noticeably stronger correlation with stellar mass or clustering near a delay time of zero (34 per cent quench within 1 Gyr of first pericentre passage) than with respect to infall (Figure~\ref{fig:QDT}).
This suggests that infall into the MW and first pericentre passage are of roughly equal importance in quenching for most satellites.
However, subsequent pericentre passages aid in quenching more massive satellites, which we show in Figure~\ref{fig:QF_peris} and Section~\ref{sec:case_studies}.
We also note that the large spread in both delay time and cosmic time at $\Mstar\lesssim10^{6.5}\,\Msun$ indicates a notably gradual transition from being quenched by reionization cutting off fresh accretion of gas \citep{Onorbe2015} to quenched by infall.

\bsp	
\label{lastpage}
\end{document}